
\magnification=1100
\hfuzz=20pt
\font\bigbf=cmbx10 scaled 1200

\centerline{\bigbf EQUILIBRIUM, STABILITY AND}
\medskip
\centerline{\bigbf ORBITAL EVOLUTION}
\medskip
\centerline{\bigbf OF CLOSE BINARY SYSTEMS}

\vskip 0.5truecm

\centerline{
Dong Lai$\,$\footnote{$^1$}{Department of Physics, Cornell University.}}
\centerline{\it Center for Radiophysics and Space Research,
Cornell University, Ithaca, NY 14853}

\bigskip
\centerline{
Frederic A.~Rasio$\,$\footnote{$^2$}{Hubble Fellow.}}
\centerline{\it Institute for Advanced Study, Princeton, NJ 08540}

\medskip
\centerline{and}

\smallskip
\centerline{
Stuart L.~Shapiro$\,$\footnote{$^3$}{Departments of Astronomy and Physics,
Cornell University.}}
\centerline{\it Center for Radiophysics and Space Research,
Cornell University, Ithaca, NY 14853}

\def\go{\mathrel{\raise.3ex\hbox{$>$}\mkern-14mu
             \lower0.6ex\hbox{$\sim$}}}
\def\lo{\mathrel{\raise.3ex\hbox{$<$}\mkern-14mu
             \lower0.6ex\hbox{$\sim$}}}
\def\eps {\varepsilon}
\def\Ich {{\cal I}}
\def\fch {{f_R}}
\def\muc {{\mu_R}}

\def\cc {{\cal C}}
\def\lone {{\lambda_1}}
\def\ltwo {{\lambda_2}}

\vskip 1.0truecm
\centerline{\bf ABSTRACT}
\vskip 0.3truecm

We present a new analytic study of the equilibrium and stability properties
of close binary systems containing polytropic components.
Our method is based on the use of ellipsoidal trial functions in an
energy variational principle.
We consider both synchronized and nonsynchronized systems,
constructing the compressible generalizations of the classical
Darwin and Darwin-Riemann configurations.
Our method can be applied to a wide variety of binary models where the
stellar masses, radii, spins, entropies, and polytropic indices are all
allowed to vary over wide ranges and independently for each component.
We find that both secular and dynamical instabilities can develop before
a Roche limit or contact is reached along a sequence of models with
decreasing binary separation. High incompressibility
always makes a given binary system more susceptible to these
instabilities, but the dependence on the mass ratio is more complicated.
As simple applications, we construct models of double
degenerate systems and of low-mass main-sequence-star binaries.
We also discuss the orbital evolution of close binary systems
under the combined influence of fluid
viscosity and secular angular momentum losses from processes
like gravitational radiation. We show that the existence of
global fluid instabilities can have a profound effect on the terminal
evolution of coalescing binaries.
The validity of our analytic solutions is examined by means of detailed
comparisons with the results of recent numerical fluid calculations
in three dimensions.

\vskip 0.5truecm
{\it Subject headings:} hydrodynamics --- instabilities --- stars:
general --- stars: binaries: close

\vskip 0.8truecm
\centerline{\bf 1. INTRODUCTION}
\vskip 0.3truecm

Essentially all recent theoretical work on close binary systems has been
done in the Roche approximation, where the noncompact components are modeled
as massless gas in hydrostatic equilibrium in the effective potential of
a point-mass system (see, e.g., Kopal 1959). This model applies well
to very compressible objects with centrally concentrated mass profiles, such as
giants and early-type main-sequence stars. Some theoretical work has
also been done in the completely opposite limit of binaries
containing a self-gravitating, {\it incompressible\/}
fluid (see Chandrasekhar 1969, and references therein; see also
Hachisu \& Eriguchi 1984b).
An essential new result found in the incompressible case is that the
hydrostatic
equilibrium solutions for sufficiently close binaries can become
{\it globally unstable\/}. The main goal of this paper is to explore
how far into the compressible domain these instabilities persist.
We parametrize compressibility by adopting a polytropic equation
of state and varying the polytropic index.

In a previous paper (Lai, Rasio, \& Shapiro 1993a, hereafter LRS1), we
have presented a comprehensive analytic study of ellipsoidal figures of
equilibrium for both single rotating polytropes and polytropes in
binary systems. We have been able to provide compressible generalizations
for all the classical incompressible solutions (Chandrasekhar 1969, hereafter
Ch69).
Our approach is based on the use of an energy variational
principle to construct approximate equilibrium solutions and study
their stability. Some applications of our results to the problem of
binary coalescence have been discussed in Lai, Rasio, \& Shapiro (1993b, c;
hereafter LRS2 and LRS3). As in Ch69, we have so far treated only
binary systems containing one star (extended component)
in orbit around a point-mass companion (as in the
classical Roche problem) or two identical stars (as in the classical Darwin
problem). We have also treated some nonsynchronized generalizations of these
cases (the so-called Roche-Riemann and Darwin-Riemann configurations;
cf.\ Aizenman 1968).

In this paper, we extend our study to the most general case of binary
systems containing two different polytropes. Specifically, we allow
the two components to have different masses, radii, entropies, polytropic
indices, and, for nonsynchronized systems, spins.
We refer to these  binary models as general Darwin-Riemann configurations.
Earlier work based on the tensor-virial method suggested that
Darwin-Riemann binary models could be constructed only for the
particular cases where the mass ratio
is either unity or tends to zero (Ch69). Instead, we show here that our
energy variational method can be generalized quite naturally to
construct binary models with arbitrary, unequal-mass components.

The usefulness of our analytic approach lies in its simplicity.
Numerical codes can be used to construct
very accurate hydrostatic equilibrium configurations in three
dimensions, but they require appreciable computer resources.
By contrast, to obtain an equilibrium model with our method simply
involves solving a set of algebraic equations, a task that can be performed
on a workstation in seconds. As a result, we
can explore a wide variety of possible binary models.
In addition, our analytic treatment can provide physical insight
into difficult issues of global stability that are easily missed
when using multidimensional numerical calculations.
In our analytic treatment, instabilities are identified simply
from turning points appearing along sequences of equilibrium configurations.
Specifically, along an equilibrium
sequence parametrized by the binary separation $r$,
the onset of instability occurs at a point $r=r_m$ where the total
energy $E$ and angular momentum $J$ of the system simultaneously
attain a {\it minimum\/}. We showed in LRS1 that such a turning point
along a binary equilibrium sequence marks the onset of instability.
Depending on the nature of the equilibrium sequence considered,
the instability can be either secular or dynamical.

As far as we are aware, {\it dynamical instabilities\/}
 of close binary systems have never
been discussed before, except in the context of the classical Darwin problem
for
incompressible fluids (Tassoul 1975; Chandrasekhar 1975).
The existence of a minimum of the total angular momentum $J$
at some $r=r_m$ has been noted before
in simple models of {\it synchronized\/} (i.e., uniformly rotating)
binary systems (Counselman 1973; Hut 1980). In synchronized systems,
the minimum comes essentially from the angular momentum and energy of the
spins, which increase as $r$ decreases.
In this case the minimum marks the onset of {\it secular instability\/}.
Counselman (1973) and Hut (1980) have discussed this
secular instability for binary systems
where the two stars can be represented by rigid spheres.
This can apply only to systems with extreme mass ratios, such as
planet-satellite systems, which have $r_m/R \gg 1$, where $R$ is the
radius of the more extended component.
In most binary systems, however, we find that $r_m/R \go 1$ and  the tidal
distortion of the more extended component cannot be neglected.

Our results have important implications for a variety of astrophysical systems
of great current interest. Both secular and dynamical instabilities
can lead to an acceleration of the orbital decay of a close binary, and,
eventually, drive the two stars to a rapid coalescence.
Close neutron star binaries are most important sources of gravitational
radiation in the Universe, and are the primary targets for the LIGO project
(Abramovici et al.\ 1992).
The coalescence of two neutron stars is at the basis of numerous models
of $\gamma$-ray bursters (see Narayan, Paczy\'nski, \& Piran 1992 and
references therein).
The consequences of fluid instabilities for the final orbital decay
of neutron star binaries and the corresponding
gravitational radiation waveforms have been explored in LRS3.
Double white dwarf systems are now generally thought to be the progenitors
of type Ia supernovae (Iben \& Tutukov 1984). They are also promising
sources of low-frequency gravitational waves
that should be easily detectable by future space-based interferometers
(Evans, Iben, \& Smarr 1987). In addition to producing supernovae, the
coalescence of
two white dwarfs may also lead in certain cases to the formation by
gravitational collapse of an isolated millisecond pulsar (Chen \& Leonard 1993)
or
the formation of  blue subdwarf stars in globular clusters (Bailyn 1993).
In the case of coalescing magnetized white dwarfs,
a neutron star with extremely high magnetic
field may form, and such an object has also been proposed as a
source of $\gamma$-ray bursts (Usov 1992).
Coalescing main-sequence star binaries are
the likely progenitors of blue stragglers
in stellar clusters (Mateo et al.\ 1990).
Contact main-sequence star binaries are also directly observed as W UMa
systems. The requirement that these systems live long enough to be observed
during a contact phase (i.e., remain stable) could place important
constraints on theoretical models of their interior structure (Rasio 1993).
Finally, mass-transfer systems, such as X-ray binaries and
cataclysmic variables, could also be affected by instabilities
if the donor star is sufficiently incompressible or massive.
A version of the secular instability described here was recently
considered by Levine et al.\ (1991, 1993) in the context of the orbital
evolution of LMC X-4 and SMC X-1.

This paper is organized as follows. In \S 2 we present our analytic method of
constructing equilibrium binary models. In \S 3 we discuss the various
stability and Roche limits for these models. We then study in \S 4
the general characteristics of
simple models for astrophysical systems containing white dwarfs,
brown dwarfs, planets, main-sequence stars, and
neutron stars. In \S 5 we compare some of our analytic results with
those of recent numerical
calculations. In \S 6 we discuss the secular orbital evolution of
close binaries in the presence of dissipation.

\vskip 0.5truecm
\centerline{\bf 2. COMPRESSIBLE DARWIN-RIEMANN MODELS}
\vskip 0.3truecm

In this section, we describe our energy variational method to construct
general Darwin-Riemann equilibrium models.
In \S 2.1 we briefly summarize the basic ideas and assumptions.
More details about the method in general, as well as many other
applications,  can be found in LRS1.
The equilibrium equations for compressible Darwin-Riemann configurations
 are derived in \S 2.2. The method of solution and the construction
of equilibrium sequences are discussed in \S 2.3.

\vskip 0.3truecm
\centerline{\bf 2.1 Basic Assumptions}
\vskip 0.2truecm

Consider an isolated, self-gravitating fluid system in steady state.
The system is characterized by conserved global quantities such as
its total mass $M$ and total angular momentum $J$.
The total energy the system (not necessarily in equilibrium) can
always be written as a functional of the fluid density and velocity
fields $\rho({\bf x})$ and ${\bf v}({\bf x})$. In principle, an
equilibrium configuration
can be determined by extremizing this energy functional with respect to all
variations of $\rho({\bf x})$ and ${\bf v}({\bf x})$ that leave the conserved
quantities unchanged. The basic idea in our method is to
replace the infinite number of degrees of freedom contained in
$\rho({\bf x})$ and ${\bf v}({\bf x})$ by a limited number of parameters
$\alpha_1,~\alpha_2,\ldots$, in such a way that the total energy
becomes a function of these parameters,
$$E=E(\alpha_1,\alpha_2,\ldots;\, M, J, \ldots),\eqno(2.1)$$
An equilibrium configuration is then
determined by extremizing the energy according to
$${\partial E \over \partial \alpha_i}=0,~~~i=1,2,\ldots\eqno(2.2)$$
where the partial derivatives are taken holding $M,\,J,\ldots\,$ constant.

An expression like~(2.1) can be written down for the total energy of
 a binary star system provided that enough simplifying assumptions are made.
In this paper, we consider only binaries in circular orbit and
we adopt a polytropic equation of state for the fluid.
Under the combined effects of centrifugal and
tidal forces, the stars assume nonspherical shapes.
We model these shapes as triaxial ellipsoids.
Specifically, we assume that the surfaces of constant density within each star
can be modeled as {\it self-similar ellipsoids\/}. The geometry is then
completely specified by the three principle axes of the outer surface.
Furthermore, we assume that the density profile $\rho(m)$ inside each star,
where $m$ is the mass interior to an isodensity surface,
is identical to that of a {\it spherical\/} polytrope with the same volume.
The velocity field of the fluid is modeled as either uniform rotation
(corresponding
to the case of a synchronized binary systems), or uniform {\it vorticity\/}
(for
nonsynchronized systems).
The vorticity vector is assumed to be everywhere parallel to the orbital
rotation
axis.

For an isolated rotating star, these  assumptions are satisfied {\it exactly\/}
when the fluid is incompressible (polytropic index $n=0$),
in which case the true equilibrium configuration is a homogeneous ellipsoid
(cf.~Ch69). For a star in a binary system, our assumptions are strictly
valid in the incompressible limit only if we truncate the tidal interaction to
quadrupole order. We adopt this quadrupole-order truncation of the
interaction potential in this paper.
For polytropes with $n\ne0$, our assumptions are only satisfied
approximately. In that case our two-ellipsoid models should be considered
as trial functions used in combination with the energy variational principle
to find approximate equilibrium solutions.

\vskip 0.3truecm
\centerline{\bf 2.2 Derivation of the Equilibrium Equations}
\vskip 0.2truecm

We consider a binary system containing two polytropes of mass $M$
and $M'$ in circular orbit around each other.
Throughout this paper unprimed quantities refer to the star of mass
$M$ and primed quantities refer to the star of mass $M'$.
Following Ch69, we denote the mass ratio as $p\equiv M/M'$.
The density and pressure are related by
$$P=K\rho^{(1+1/n)},~~~~P'=K'\rho'^{\,(1+1/n')}.\eqno(2.3)$$
Note that we allow the two stars to have different
polytropic indices ($n\ne n'$) and different polytropic constants ($K\ne K'$).
This allows us to model realistically a variety of astrophysical systems
where the two stars have different masses and radii
(see \S 4). The binary separation is denoted by $r$,
and the principal axes of the two ellipsoids by $a_1$, $a_2$, $a_3$, and
$a_1'$, $a_2'$, $a_3'$. The orientation is such that
$a_1$ and $a_1'$ are measured along the binary axis,
$a_2$ and $a_2'$ in the direction of the orbital motion, and $a_3$ and
$a_3'$ along the rotation axis.
In place of the three principle axes $a_i$, it is often convenient to introduce
as independent variables
the central density $\rho_c$, and two oblateness parameters defined as
$$\lambda_1 \equiv \biggl({a_3\over a_1}\biggr)^{2/3},~~~~
\lambda_2 \equiv \biggl({a_3\over a_2}\biggr)^{2/3}.
\eqno(2.4)$$
Similarly we can introduce $\rho_c'$, $\lambda_1'$ and $\lambda_2'$
in place of the three $a'_i$.
Thus the seven independent variables which specify the structure of
our models are
$\{\alpha_i,\,\, i=1,\cdots,7\}=\{r,\,\,\rho_c,\,\lambda_1,\,\lambda_2,
\,\,\rho_c',\,\lambda_1',\,\lambda_2'\}$.

\bigskip
\centerline{\it 2.2.1 Energy Function}
\medskip

We first obtain an expression for the total energy of the system
under the assumptions given in \S 2.1. A detailed derivation
for the similar case of Roche-Riemann configurations was given
in LRS1 and will not be repeated here.
 References to key equations
in LRS1 are  indicated with numbers preceded by an ``I''.
Throughout this section, when two similar expressions can be written for
the two stars, we only give the one corresponding to $M$, the other being
obtained trivially by replacing unprimed by primed quantities.

The total internal energy in each star is given by (cf.\ eq.~[I.3.1])
$$U=\int\! {nP\over \rho}\, dm = k_1K\rho_c^{1/n}M.\eqno(2.5)$$
The self-gravitational potential energy can be written (cf.\ eq.~[I.4.6])
$$W=-{3\over5-n}{GM^2\over R}f
=- k_2GM^{5/3}\rho_c^{1/3} f,\eqno(2.6)$$
where we have defined the mean radius $R\equiv (a_1a_2a_3)^{1/3}$
and the dimensionless ratio
$$f=f(\lambda_1,\lambda_2)\equiv {A_1a_1^2+A_2a_2^2+A_3a_3^2 \over 2 R^2},
\eqno(2.7)$$
such that $f=1$ for a spherical star.
The index symbols $A_i$ are defined as in Ch69 (\S17),
$$A_i\equiv a_1a_2a_3\int_0^{\infty}\!\!{du \over \Delta (a_i^2+u)},~~~~
{\rm with}~~~~\Delta^2=(a_1^2+u)(a_2^2+u)(a_3^2+u).\eqno(2.8)$$
They are functions of $\lambda_1$ and $\lambda_2$ only.
In equations~(2.5) and~(2.6), $k_1$ and $k_2$ are dimensionless polytropic
structure constants (depending only on $n$), defined as
$$k_1\equiv{n(n+1)\over5-n}\,\xi_1|{\theta'}_1|,~~~~
k_2\equiv{3\over5-n}\,\biggl({4 \pi |{\theta'}_1|\over\xi_1}\biggr)^{1/3},
\eqno(2.9)$$
where $\theta$ and $\xi$ are the usual Lane-Emden variables for a polytrope
(see, e.g., Chandrasekhar 1939). Values of $k_1$ and $k_2$ for different
$n$ are given in Table~1 of LRS1, but they are not needed explicitly for
constructing equilibrium solutions.

The fluid velocity field inside each star is modeled exactly as in
a Riemann-S ellipsoid (see LRS1, \S5, for details).
Following Ch69 and LRS1, we introduce a parameter $\fch$ defined as
$$\fch\equiv{\zeta\over\Omega},\eqno(2.10)$$
where $\Omega$ is the orbital angular frequency and
$\zeta$ is the fluid vorticity in the corotating frame,
$$\zeta\equiv(\nabla\times {\bf u})\cdot {\bf e}_3
=-{a_1^2+a_2^2\over a_1a_2}\Lambda.\eqno(2.11)$$
The quantity $\Lambda$ is the angular frequency of the internal fluid motions.
The velocity field in the corotating frame is given by
$${\bf u}=Q_1 x_2 {\bf e}_1+Q_2 x_1 {\bf e}_2,\eqno(2.12)$$
where
$$\eqalign{
  Q_1&=-{a_1^2\over a_1^2+a_2^2}\,\zeta=+{a_1\over a_2}\Lambda, \cr
  Q_2&=+{a_2^2\over a_1^2+a_2^2}\,\zeta=-{a_2\over a_1}\Lambda. \cr
}\eqno(2.13)$$
Here $\bf e_1$ is along the binary axis, directed from $M$ to $M'$,
$\bf e_2$ is in the direction of the orbital velocity, and the origin
is at the center of mass of $M$.
The fluid velocity in the inertial frame is given by
$${\bf u}^{(0)}={\bf u}+{\bf\Omega}\times {\bf x},\eqno(2.14)$$
and the vorticity in the inertial frame is
$${\zeta}^{(0)}\equiv(\nabla\times{\bf u}^{(0)})\cdot {\bf e}_3
=(2+\fch)\,\Omega.\eqno(2.15)$$
Uniform rotation (synchronization) corresponds to $\fch=\zeta=\Lambda=0$.
An {\it irrotational\/} velocity field is obtained when $\fch=-2$.
Note that the two equilibrium figures (the geometric outer shapes
of the two stars)  always rotate at the orbital angular
frequency $\Omega$ as seen in the inertial frame.

An expression for the ``spin'' kinetic energy $T_s$
(i.e., the kinetic energy in internal fluid motions)
in the inertial frame can be obtained from equations~(2.12)--(2.14).
One finds (cf.\ eq.~[I.5.6])
$$T_s={1\over2}I(\Lambda^2+\Omega^2)
-{2\over5}\kappa_nMa_1a_2\Lambda\Omega,\eqno(2.16)$$
where
$$I={1\over 5}\kappa_n M (a_1^2+a_2^2)\eqno(2.17)$$
is the moment of inertia. We have defined the dimensionless coefficient
$$\kappa_n\equiv{5\over3}\,{\int_0^{\xi_1}\theta^n\xi^4\,d\xi\over
\xi_1^4|{\theta'}_1|},\eqno(2.18)$$
so that $\kappa_n=1$ for $n=0$.
Values of $\kappa_n$ are given in Table~1 of LRS1.
Similarly, the ``spin'' angular momentum $J_s$ can be written (cf.\ eq.~[I.5.5)
$$J_s =I\Omega -{2\over5}\kappa_nMa_1a_2\Lambda.\eqno(2.19)$$
Another important conserved quantity is the fluid circulation $C$
along the equator of the star. Following LRS1 (\S 5.1) we write
$$\cc\equiv \biggl (-{1\over5\pi}\kappa_nM\biggr)\,C
=\biggl (-{1\over5\pi}\kappa_nM\biggr)\,\pi a_1a_2\, {\zeta}^{(0)}
= I\Lambda-{2\over5}\kappa_nMa_1a_2\Omega.\eqno(2.20)$$
The quantity $\cc$ has the dimensions of angular momentum but is proportional
to the conserved circulation $C\equiv \pi a_1a_2 {\zeta}^{(0)}$.
We usually refer to $\cc$ itself as the circulation for convenience.
Note the complementary roles played
by the variables $(J_s,\cc)$ and $(\Omega,\Lambda)$
(compare expressions~[2.19] and~[2.20]).

We can now obtain simple expressions for the total kinetic energy and
angular momentum in the system.
We first rewrite $T_s$ in a form that is more convenient for taking
derivatives of the energy function (cf.\ LRS1, \S 5.1).
We first combine equations~(2.19) and~(2.20) to derive the result
$$J_s \pm \cc ={1\over 2}I_{\pm}(\Omega \pm \Lambda),\eqno(2.21)$$
where
$$I_{\pm}\equiv{2\over 5} \kappa_n M(a_1 \mp a_2)^2 = 2 I_s/h_{\pm}.
\eqno(2.22)$$
Here $I_s \equiv{2\over 5} \kappa_n M R^2$ is the moment of inertia of a sphere
with
the same volume as the ellipsoid, and $h_{\pm}\equiv 2R^2/(a_1 \mp a_2)^2$.
Using equation~(2.21), the kinetic energy $T_s$ can then be expressed as
$$T_s=T_{+}+T_{-},\eqno(2.23)$$
with
$$T_{\pm}={1\over 8}I_{\pm}(\Omega \pm \Lambda)^2
={1\over 2I_{\pm}}(J_s \pm \cc)^2.
\eqno(2.24)$$
The orbital contributions to the total angular momentum and kinetic energy
are simply
$$J_o=\mu r^2\Omega,~~~~T_o={1\over 2}\mu \Omega^2 r^2,\eqno(2.25)$$
where $\mu=MM'/(M+M')$ is the reduced mass
\footnote{$^4$}{Note that the quantity $\mu\equiv GM'/r^3$ introduced
in LRS1 is called $\mu_R$ in this paper; see below.}.
The total angular momentum and kinetic energy of
the system are then given by
$$\eqalign{
J& =J_s+J_s'+J_o,\cr
T& =T_s+T_s'+T_o.\cr
}\eqno(2.26)$$

Finally, the gravitational interaction energy $W_i$ between the two stars
is given, up to quadrupole order, by (see Appendix~B of LRS1)
$$W_i=-{GMM'\over r}-{GM\over 2r^3}(2I_{11}'-I_{22}'-I_{33}')
-{GM'\over 2r^3}(2I_{11}-I_{22}-I_{33}),\eqno(2.27)$$
where we have defined
$$I_{ij}={1\over 5}\kappa_n M a_i^2\delta_{ij},~~~~~
I_{ij}'={1\over 5}\kappa_n' M' a_i'^2\delta_{ij}.
\eqno(2.28)$$

The total energy of the system, not necessary in equilibrium, is given by
$$E(r,\rho_c,\lambda_1,\lambda_2,
\rho_c',\lambda_1',\lambda_2';\,M,M',\cc,\cc',J)
=U+U'+W+W'+T+W_i,\eqno(2.29)$$
together with equations~(2.5), (2.6), and (2.23)--(2.27).

When extremizing the energy function~(2.29) to find equilibrium solutions,
we must hold all the conserved quantities $M$, $M'$, $J$, $\cc$, and $\cc'$
 constant. Clearly, the form of $T$ given by
equations (2.23)--(2.26) is not ideally suited to such a procedure, since
its dependence on $J,~\cc$, and $\cc'$ is not explicit.
However, using equations~(2.21) and~(2.26), we can express $\Omega$ as a
function of the adopted variables $\{\alpha_i\}$ and the conserved quantities
as
$$\Omega={1\over \mu r^2+I_A}
\biggl [J+\cc\biggl({h_{+}-h_{-}\over h_{+}+h_{-}}\biggl)
+\cc'\biggl({h_{+}'-h_{-}'\over h_{+}'+h_{-}'}\biggl)\biggl],\eqno(2.30)$$
where
$$I_A\equiv {2I_s\over h_{+}+h_{-}}+{2I_s'\over h_{+}'+h_{-}'}.\eqno(2.31)$$
Expressions for $\Lambda$ and $\Lambda'$ can then be obtained using
equation~(2.20). In principle, we can substitute these expressions
into equation~(2.29) and obtain an appropriate expression for $T$
which does not contain $\Omega,~\Lambda$ or $\Lambda'$,
and depends only on $\{\alpha_i\}$ and the conserved quantities.
Instead, it is more convenient to use the following expression for $T$,
$$T={(J_s+ \cc)^2\over 2I_{+}} + {(J_s-\cc)^2\over 2I_{-}}
+{(J-J_s-\mu r^2\Omega+\cc')^2\over 2I_{+}'}
+{(J-J_s-\mu r^2\Omega-\cc')^2\over 2I_{-}'}
+{1\over 2}\mu r^2\Omega^2.\eqno(2.32)$$
Although $\Omega$ and $J_s$ appear in this expression, we can
treat them as constant parameters when taking a
first derivative of $E$ with respect to $\alpha_i$. This is because
$\partial E/\partial\Omega=\partial T/\partial\Omega=0$ and
$\partial E/\partial J_s=0$, as can be shown easily
with the help of equation~(2.21).

\bigskip
\centerline{\it 2.2.2 Equilibrium Conditions}
\medskip

We can now derive the equilibrium conditions~(2.2) for a
general Darwin-Riemann configuration.

The first equilibrium condition, $\partial E/\partial r=0$, gives a
relation between $\Omega^2$ and $r$, i.e.,
the {\it modified Kepler's law\/} for the binary (cf.\ eq.~[I.7.6]),
$$\eqalign{
\Omega^2 &= {G(M+M')\over r^3}\,(1+\delta+\delta')\cr
& =\muc (1+p) (1+\delta+\delta')
=\muc' (1+{1\over p}) (1+\delta+\delta'),\cr
}\eqno(2.33)$$
where we have defined
$$\muc \equiv GM'/r^3,~~~~\muc' \equiv GM/r^3,\eqno(2.34)$$
and
$$\delta\equiv{3\over2}\,{(2I_{11}-I_{22}-I_{33})\over Mr^2},~~~~
\delta'\equiv{3\over2}\,{(2I_{11}'-I_{22}'-I_{33}')\over M'r^2}.
\eqno(2.35)$$
Note that for Roche and (equal-mass) Darwin configurations, Ch69 uses
$\Omega^2=\Omega_K^2=G(M+M')/r^3$ and $\Omega^2=\Omega_K^2(1+\delta)$,
respectively, whereas our equation~(2.33) gives $\Omega^2=\Omega_K^2(1+\delta)$
and $\Omega^2=\Omega_K^2(1+2\delta)$. Our value of $\Omega^2$ is more
accurate than that used by Ch69 (see LRS1, Appendix~C, for a complete
discussion). Moreover, in our treatment, there is no restriction on the
allowed ratio of $M/M'$.

The second condition, $\partial E/\partial \rho_c=0$, leads
to the {\it  virial relation\/} for the star of mass $M$ (cf.\ eq.~[I.7.8]),
$${3\over n}U+W+2T_s=-{GMM'\over R}g_t,\eqno(2.36)$$
where
$$g_t\equiv {R\over Mr^3}\,(2I_{11}-I_{22}-I_{33})
={2\over3}\,{R\over r} \delta.\eqno(2.37)$$
{}From equations (2.5), (2.6), and~(2.36), the equilibrium mean radius can be
obtained as
$$R=R_o\biggl [f(\lone,\ltwo)\biggl(1-2{T_s\over|W|}\biggr)
-\biggl ({5-n \over 3p}\biggr)\,g_t\biggr]^{-n/(3-n)},\eqno(2.38)$$
where $R_o$ is the radius of the spherical polytrope,
$$R_o=\xi_1 \left(\xi_1^2|\theta'_1|\right)^{-(1-n)/(3-n)}
\biggl [{(n+1)K\over4\pi G}\biggr]^{n/(3-n)}
\biggl ({M\over4\pi}\biggr)^{(1-n)/(3-n)}.\eqno(2.39)$$
Clearly, the third condition $\partial E/\partial \rho_c'=0$
gives  a similar expression for $R'$,
$$R'=R_o'\biggl [f(\lone',\ltwo')\biggl(1-2{T_s'\over|W'|}\biggr)
-\biggl ({5-n' \over 3}\biggr)\,p g_t'\biggr]^{-n'/(3-n')}.\eqno(2.40)$$

The fourth and fifth conditions,
$\partial E/\partial\lone=\partial E/\partial\ltwo=0$,
together with the virial relation~(2.36) can be written in
the form (cf.\ eq.~[I.8.4])
$$\eqalign{
-{U\over n}&={\cal M}_{11}+I_{11}(\Omega^2+2\muc+2\Omega Q_2)+I_{22}Q_1^2,\cr
-{U\over n}&={\cal M}_{22}+I_{22}(\Omega^2- \muc-2\Omega Q_1)+I_{11}Q_2^2,\cr
-{U\over n}&={\cal M}_{33}-\muc I_{33},\cr
}\eqno(2.41)$$
where the $Q_i$ are given in equation~(2.13). Here
we have introduced the potential-energy tensor (Ch69)
$${\cal M}_{ij}\equiv-2\pi G{\bar\rho}A_i\left(
    {Ma_i^2\delta_{ij}\over5-n}\right)~~~~~~~~{\rm(no~summation)},\eqno(2.42)$$
such that $W={\cal M}_{ii}$. The quantity
$\bar\rho \equiv 3M/(4\pi R^3)$ is the mean density.
The form of equations~(2.41) is identical to that of
the corresponding equations for Roche-Riemann configurations
(LRS1, \S8), but the expression for $\Omega$ is different.
Equations~(2.41) and~(2.42) can be used to obtain relations for the principal
axes (cf.\ eqs.~[I.8.5], [I.8.6]),
$$\eqalign{
q_n\,{\tilde \mu}_R
\biggl\lbrace {Q_{1}^2\over\muc}a_2^2
+ \biggl [2+(1+p)(1+\delta+\delta')+2{Q_{2}\Omega\over\muc}\biggr]\,a_1^2
+a_3^2\biggr\rbrace
&=2 (a_1^2A_1-a_3^2A_3),\cr
q_n\,{\tilde \mu}_R
\biggl\lbrace {Q_{2}^2\over\muc}a_1^2
+\biggl [(1+p)(1+\delta+\delta')-1-{2Q_{1}\Omega\over\muc}\biggr]\,a_2^2
+a_3^2\biggr\rbrace
&=2(a_2^2A_2-a_3^2A_3),\cr
}\eqno(2.43)$$
where we have defined
$$q_n\equiv \kappa_n \biggl(1-{n\over 5}\biggr),~~~~
{\tilde \mu}_R \equiv {\muc \over \pi G\bar\rho}.\eqno(2.44)
$$
Using equations~(2.13) and~(2.33) we see that
the quantities $Q_i^2/\mu_R$ and $Q_i\Omega/\mu_R$ appearing in
equations~(2.43) are given by
$${Q_{i}^2\over\muc}={\hat Q}_i^2(1+p)(1+\delta+\delta')\fch^2,~~~~
{Q_{i}\Omega\over\muc}={\hat Q}_i (1+p)(1+\delta+\delta')\fch,\eqno(2.45)$$
where
$${\hat Q}_1=-{a_1^2\over a_1^2+a_2^2},~~~~~
{\hat Q}_2={a_2^2\over a_1^2+a_2^2}.\eqno(2.46)$$
Here again, the two structure equations~(2.43) are very similar to
those obtained in the Roche-Riemann case.
The similarity results from the fact that, to quadrupole order,
each star acts on the other like a point mass. The only real coupling
is through $\Omega$: in equations~(2.43) and~(2.45) we have the factor
$(1+\delta+\delta')$ instead of simply $(1+\delta)$ in the Roche-Riemann case.

Finally, the last two conditions,
$\partial E/\partial\lone'=\partial E/\partial\ltwo'=0$,
yield two structure equations similar to equations~(2.43)
for the other component
$$\eqalign{
q_n'\,{\tilde \mu}_R'\biggl\lbrace {Q_{1}'^2\over\muc'}a_2'^2
+ \biggl [2+(1+{1\over p})(1+\delta+\delta')
+2{Q_{2}'\Omega\over\muc'}\biggr]\,a_1'^2
+a_3'^2\biggr\rbrace
&=2 (a_1'^2A_1'-a_3'^2A_3'),\cr
q_n'\,{\tilde \mu}_R'\biggl\lbrace {Q_{2}'^2\over\muc'}a_1'^2
+\biggl [(1+{1\over p})(1+\delta+\delta')-1
-{2Q_{1}'\Omega\over\muc'}\biggr]\,a_2'^2
+a_3'^2\biggr\rbrace
&=2(a_2'^2A_2'-a_3'^2A_3'),\cr
}\eqno(2.47)$$
where $q_n'\equiv \kappa_n'(1-n'/5)$,
${\tilde \muc}' \equiv \muc'/(\pi G\bar \rho')$, and
$\bar \rho'\equiv 3M'/(4\pi R'^3)$.

The seven equations in~(2.33), (2.38), (2.40), (2.43), and~(2.47)
completely determine the equilibrium configurations.
Once the equilibrium values of the seven variables $\{\alpha_i\}$ are
determined,
the total equilibrium
angular momentum can be obtained from equations~(2.19), (2.25) and~(2.26).
Using the virial relations (eq.~[2.36] and the corresponding expression
for $M'$), the total equilibrium energy of the system can be written
explicitly
$$\eqalign{
E_{eq}&=E_s+E_s'+{1\over 2}\mu r^2\Omega^2-{GMM'\over r}\cr
&-\biggl ({2 n+3\over 6}\biggr){GM'\over r^3}(2I_{11}-I_{22}-I_{33})
-\biggl ({2 n'+3\over 6}\biggr){GM\over r^3}(2I_{11}'-I_{22}'-I_{33}'),\cr
}\eqno(2.48)$$
where
$$E_s=-{(3-n)GM^2 \over (5-n)R}f(\lone,\ltwo)
\biggl[1-\biggl({3-2n\over3-n}\biggr)\,
{T_s\over |W|}\biggr],\eqno(2.49)$$
and the expression for $E_s'$ is similar.

\vskip 0.3truecm
\centerline{\bf 2.3 Construction of Equilibrium Sequences}
\vskip 0.2truecm

For many astrophysical applications, it is useful to construct
sequences of equilibrium configurations with varying binary
separation. The onset of instabilities can often be determined by
locating turning points along such sequences (see \S 3). In addition,
equilibrium sequences can sometimes be used to describe approximately
the orbital evolution of a system driven by some dissipation
mechanism (see \S 6).
Different types of equilibrium sequences can be constructed
depending on which quantities are held constant along the sequence.

A particularly useful dimensionless ratio that can be used to parametrize
an equilibrium sequence is
$$\hat r\equiv {r\over a_1+a_1'}. \eqno(2.50)$$
Because $a_1$ and $a_1'$ can be double-valued functions of $r$ in our
models, the quantity $\hat r$ does {\it not\/}, in general, specify uniquely
the absolute separation between the centers of mass of the two components.
Instead, it is a measure of how close the surfaces of the two
ellipsoids are approaching each other.
The usefulness of the definition~(2.50) in practice lies in the existence of
a {\it unique\/} equilibrium solution for each value of $\hat r$
along the sequence.
In particular, the {\it contact solution\/}, corresponding to the point along a
sequence where the surfaces of two ellipsoids are first touching,
 can be readily determined by setting $\hat r=1$.
Physical solutions require $\hat r\ge 1$ in our models, since the
energy function is calculated for two ellipsoids that do not overlap.
Therefore, all our equilibrium sequences terminate at the contact solution.

\bigskip
\centerline{\it 2.3.1 Darwin-Riemann Sequences}
\medskip

Following traditional practice for sequences of classical Riemann-S
and Roche-Riemann ellipsoids (Ch69), we define a Darwin-Riemann sequence as
a sequence of Darwin-Riemann configurations along which the quantities
$f_R$ and $f'_R$ (cf.\ eq.~[2.10]) are held constant.

Darwin-Riemann sequences are constructed as follows. We adopt units such that
$G=M=R_o=1$ and fix the values of $p=M/M'$, $R_o/R'_o$ (or,
equivalently, $K/K'$), $f_R$, and $f'_R$ along the sequence.
For each value of $1\le{\hat r}<\infty$, we solve the
equilibrium equations for the variables
$(\hat x_j)=(\hat a_1,\hat a_2,\hat a_3,\hat a_1',\hat a_2',\hat a_3')$,
where $\hat a_i\equiv a_i/(a_1+a_1')$ and $\hat a_i'\equiv a_i'/(a_1+a_1')$.
Specifically, we solve the coupled set of algebraic equations numerically by a
Newton-Raphson iteration method. The following six functions must vanish
simultaneously:
$$\eqalign{
F_1 &=\hat a_1+\hat a_1'-1,\cr
F_2 &={R\over R'}-\biggl({a_1a_2a_3\over a_1'a_2'a_3'}\biggl)^{1/3},~~~~
{\rm with~}R,\,R'~{\rm given~by~eqs.~[2.38]~and~[2.40]},\cr
F_3 &={\rm LHS (eq.~[2.43a]) - RHS (eq.~[2.43a])},\cr
F_4 &={\rm LHS (eq.~[2.43b]) - RHS (eq.~[2.43b])},\cr
F_5 &={\rm LHS (eq.~[2.47a]) - RHS (eq.~[2.47a])},\cr
F_6 &={\rm LHS (eq.~[2.47b]) - RHS (eq.~[2.47b])}.\cr
}\eqno(2.51)$$
We use the standard Newton-Raphson algorithm described by
Press et al.\ (1987), except that here the matrix
$(\partial F_i/\partial \hat x_j)$ needed in every
Newton-Raphson iteration must be evaluated numerically by finite difference.
An equilibrium solution can be calculated
to an accuracy of $\delta a_i/a_i < 10^{-5}$
in $\sim 1$ CPU second on a Sun SPARC workstation.
We start the calculation for a large value of $\hat r$, so that
spherical stars can be used as an initial guess.
As $\hat r$ is decreased, we use as initial guess the
equilibrium solution previously determined for a slightly larger $\hat r$.
Once the principal axes are determined, other physical quantities
such as $E$, $J$, and $\Omega$
can then be calculated using the expressions given in \S 2.2.

{}From equation (2.29), we can easily prove that the relation
$$dE=\Omega \,dJ+\Lambda \,d\cc +\Lambda'\,d\cc',\eqno(2.52)$$
(cf.\ LRS1, Appendix~D).
must be satisfied along any equilibrium sequence.
This provides a convenient check on our numerical results.
In particular, for synchronized sequences ($\Lambda=\Lambda'=0$) and
constant-circulation sequences (see below), we must have $dE=\Omega\, dJ$
(cf.\ Ostriker \& Gunn 1969).

\bigskip
\centerline{\it 2.3.2 Compressible Darwin Sequences}
\medskip

The classical Darwin problem (Ch69) concerns two identical, incompressible
objects ($n=0$) in a {\it synchronized\/} (uniformly rotating) system.
Here we generalize the classical Darwin configurations by allowing
for compressibility as well as nonidentical components.
The synchronized Darwin sequences can be constructed as a special case
of the general Darwin-Riemann sequences where we set $\fch=\fch'=0$.
In this case we have
$\Lambda=\Lambda'=0$ and there are no internal fluid motions
in the corotating frame of the binary. As an illustration,
Figure~1 shows the variation of the total equilibrium energy $E$
(eq.~[2.48]), total angular momentum $J$, and orbital frequency $\Omega$
along selected Darwin sequences.
Note that the kinetic energy (eq.~[2.32]) for Darwin configurations can be
written simply
$$T={1\over 2}(\mu r^2+I+I')\,\Omega^2
={J^2 \over 2\,(\mu r^2+I+I')}.\eqno(2.53)$$

Synchronization can be achieved in close binaries if there is a large
enough effective viscosity to maintain the tidal coupling between
the spins and the orbital motion. This is expected to be the case
 for the majority of close binaries with the possible exception
of double neutron star systems (see \S 4.3).
If there is another dissipation mechanism, e.g., gravitational radiation,
that drives orbital decay on a time scale much longer than the viscous
dissipation time scale, then we can expect
the system to evolve along a synchronized equilibrium sequence (see \S 6).

\bigskip
\centerline{\it 2.3.3 Constant-Circulation Sequences}
\medskip

In the opposite limit, when viscosity is completely negligible,
a systen whose orbit is decaying because of gravitational radiation
evolves along a sequence of configurations with constant $\cc$ and $\cc'$.
This is because the gravitational radiation reaction forces conserve the
fluid circulation (Miller 1974). Constant-circulation sequences may
describe the orbital evolution of coalescing neutron star binaries (Kochanek
1992).

The value of $\cc$ is set by the spin
angular frequency $\Omega_s$ of $M$ when the binary separation is large;
similarly for $\cc'$.
{}From equation~(2.33) we see that, for large $r$, we may set
$\Omega^2=G(M+M')/r^3$, and we then have (cf.\ eqs.~[2.19] and~[2.20])
$$\eqalign{
J & \rightarrow  \mu r^2 \Omega - I \Lambda-I' \Lambda'
= \mu r^2 \Omega + I \Omega_s+I'\Omega_s',\cr
\cc &\rightarrow I\Lambda= -I\Omega_s,\cr
\cc'&\rightarrow I'\Lambda' = -I'\Omega_s',
}\eqno(2.54)$$
where we have identified $\Omega_s=-\Lambda(r=\infty)$ and
$\Omega_s'=-\Lambda'(r=\infty)$. Note that
when $\Omega_s$ is positive (i.e., the spin is in the same direction
as the orbital $\Omega$), $\cc$ is negative.

A particularly interesting special case is the so-called {\it irrotational\/}
Darwin-Riemann sequence, for which $\cc=\cc'=0$.
{}From equation~(2.15), we see that $\fch=\fch'=-2$ are also constant
along such a
sequence. This corresponds to the case where the stars have negligible
spin at large $r$ and evolve in the absence viscosity.

In general $\fch$ and $\fch'$  vary along the constant-circulation
sequences. Numerically, for given
$\hat r$, eight variables are now required to specify an equilibrium solution:
the six variables $\hat x_j$ introduced previously plus $\fch$ and $\fch'$.
Two more functions need to be set equal to zero in the Newton-Raphson scheme
(cf.~eqs.~[2.15] and [2.20]):
$$\eqalign{
F_7&=\biggl (-{1\over5\pi}\kappa_nM\biggr)\pi a_1a_2(2+\fch)\Omega-\cc,\cr
F_8&=\biggl (-{1\over5\pi}\kappa_n'M'\biggr)\pi a_1'a_2'(2+\fch')
\Omega-\cc'.\cr
}\eqno(2.55)$$

\bigskip
\centerline{\it 2.3.4 Sequences with Constant Angular Momentum}
\medskip

Since viscous dissipation conserves the total angular momentum,
a binary system evolving through viscosity only
will follow a sequence of configurations with constant $J$.
Note that, for a given value of $J$, the equilibrium sequence is not
uniquely determined, since
the two stars can have different spins to give the same total
angular momentum. However, if initial values of $\cc$ and $\cc'$ (or $f_R$
and $f_R'$) are specified, then a unique sequence with $J=\,$constant
starting from those initial values can be constructed,
provided that we know the energy dissipation rate due to viscosity.
Constant-$J$ sequences are discussed in more details in \S 6.2.2.

\bigskip
\centerline{\it 2.3.5 Asymptotic Solutions for Large $r$}
\medskip

Explicit solutions of our equations can be obtained analytically
in the limit where $r\gg R_o+R_o'$. The results are summarized
in Appendix~A for several types of equilibrium sequences.
These asymptotic solutions for large $r$ have been used
in LRS3 to calculate the lowest-order deviations from the point-mass
results for the gravitational radiation waveforms emitted during the
coalescence of two neutron stars.

\vskip 0.5truecm
\centerline{\bf 3. STABILITY LIMITS AND ROCHE LIMITS}
\vskip 0.3truecm

\vskip 0.3truecm
\centerline{\bf 3.1 Secular and Dynamical Stability Limits}
\vskip 0.2truecm

In general, stability requires that an equilibrium configuration correspond to
a true {\it minimum\/} of the total energy
$E(\alpha_1,\alpha_2,\ldots; M,J,\ldots)$, i.e., that all eigenvalues of the
matrix $(\partial^2 E/\partial \alpha_i\partial \alpha_j)_{eq}$
be positive. The onset of instability along any one-parameter sequence of
equilibrium configurations can be determined from the condition (LRS1, \S2.3)
$${\rm det}\biggl({\partial^2E\over\partial \alpha_i\partial \alpha_j}
\biggr)_{eq}=0,~~~~i,j=1,2,\ldots~~~~~~({\rm onset~of~instability}).
\eqno(3.1)$$
When this condition is satisfied, one of the eigenvalues must change
sign. It may then become possible for the system to further minimize
its energy by evolving away from the equilibrium configuration considered.

As discussed in LRS1 (\S2.4), the nature of the instability
depends on the type of perturbation considered about equilibrium.
A {\it dynamical\/} instability can develop only
when $J,~\cc$ and $\cc'$ are all conserved by the perturbation;
a {\it secular\/} instability requires viscous dissipation, which conserves
 only $J$.
Mathematically, we locate the point of onset of dynamical instability
by evaluating the second derivatives
in equation~(3.1) holding $J,~\cc$ and $\cc'$ all fixed.
If instead we evaluate
equation~(3.1) by fixing $J$ alone, then we locate the point of onset
of secular instability.
For general Darwin-Riemann configurations,
there are 28 independent matrix elements that need to be
evaluated in equation~(3.1). They are listed in Appendix~B.

Alternatively, we can show that the stability limits determined from
equation~(3.1) coincide with {\it turning points\/} along
appropriately constructed equilibrium sequences (LRS1, \S2.3). Specifically,
a dynamical stability limit coincides with the point where
the total equilibrium energy and angular momentum are both minimum
along a constant-circulation sequence. A secular stability
limit coincides with similar minima appearing along a synchronized sequence.
Note that the minima in $E$ and $J$ must coincide along
synchronized or constant-circulation sequences since
$dE=\Omega\, dJ$ along such sequences (cf.\ eq.~[2.52] and Appendix~D of LRS1).

Nonsynchronized configurations can never be true equilibria in the
presence of viscosity.
Therefore we consider only {\it dynamical\/}
 stability along constant-circulation sequences.
A minimum of $E$ and $J$ at some $r=r_m$
along these sequences indicates the onset of dynamical instability.
Indeed, at $r=r_m$, it becomes possible for
a small dynamical perturbation of the system (which conserves
$\cc$ and $\cc'$) to cause no first-order change in the total equilibrium
energy or angular momentum. This indicates a change of sign in the
eigenfrequency $\omega^2$ associated with the perturbation (see, e.g.,
Shapiro \& Teukolsky 1983, Chap.~6).

Along synchronized equilibrium sequences, both secular and dynamical
stability limits can exist.
A minimum of $E$ and $J$ along these sequences marks the onset of
{\it secular\/} instability. In the presence of
viscosity, configurations with $r<r_m$ will be driven away from
synchronization in order to minimize their energy (see \S 6).
The instability at $r=r_m$ cannot be dynamical here
because the neighboring configurations along the sequence are still
in uniform rotation and cannot be reached by a small perturbation
unless viscosity is present. True dynamical instability along a
synchronized sequence can occur at some  $r<r_m$
where neighboring configurations {\it with the same values of\/}
$\cc$ and $\cc'$ can be reached with no change in total equilibrium
energy to first order. In practice, to obtain the
dynamical stability limit along a synchronized sequence
parametrized by $\hat r$, we can proceed as follows.
At every $\hat r$, we calculate the equilibrium energy of the synchronized
configuration $E_{eq}(\hat r)$ and the corresponding values of $\cc,~\cc'$.
We then construct a neighboring equilibrium model, with $\hat r$ larger by a
small
increment $\Delta \hat r$, having the same values of $\cc$ and $\cc'$,
and obtain its energy $\tilde E_{eq}({\hat r}+\Delta{\hat r})$. The onset of
 dynamical instability is then located where
$({\tilde E_{eq}}({\hat r}+\Delta{\hat r})-E_{eq}(\hat r))/\Delta{\hat r}=0$.

We can use either the determinant equation~(3.1) or the turning point method
to locate the critical instability points along an equilibrium sequence.
The two methods are mathematically equivalent (cf.~LRS1, \S2.3).
In many cases we have used both criteria to verify the numerical accuracy of
our
identifications.

To illustrate these points, consider an example where the binary contains
two identical components with $n=0$ (incompressible fluid).
We show in Figure~2 how the total equilibrium energy $E_{eq}$ varies along the
synchronized sequence, as well as several sequences with constant circulation.
The minimum of $E_{eq}(r)$ along the synchronized sequence
indicates the onset of secular instability, while the minima
along the constant-circulation curves correspond to the onset of
dynamical instability. We see that there is a unique point on the
synchronized sequence at which the constant-circulation curve that intersects
it
attains a minimum. This point corresponds to the
onset of dynamical instability along the synchronized sequence.

\vskip 0.3truecm
\centerline{\bf 3.2 Roche Limits and Contact Configurations}
\vskip 0.2truecm

When the masses of the two binary components are different,
a {\it Roche limit\/} may exist prior to contact along an equilibrium sequence.
Recall that we define contact simply as the point where the surfaces
of the two ellipsoids first touch.
The Roche limit in our ellipsoidal models is defined as the point where
the binary separation $r$ has a minimum value below which
no equilibrium solution exists. At the Roche limit, the slope of the
$E_{eq}(r)$ curve becomes infinite (cf.\ Fig.~1).
This behavior is somewhat artificial, resulting probably from the
truncation of the interaction potential to quadrupole order.
The same effect can be seen, but less marked,
even in fully numerical solutions going to much higher order
but still retaining only a finite number of terms in the multipole
expansion of the interaction potential (see \S5).

Equilibrium solutions continue to exist beyond the Roche limit
(at smaller $\hat r$) in some of our models. This second branch of
solutions beyond the Roche limit is clearly unphysical since it
has higher energy than the main equilibrium branch for the
same value of $J$. The equilibrium configurations beyond the Roche limit
 must therefore be at least secularly unstable (see \S 6).
Indeed, we note that the Roche limit itself must already be
situated beyond the point where $E$ and $J$ are minimum,
indicating instability.

Equilibrium sequences for systems with a mass ratio $p=M/M'$ sufficiently
close to unity may terminate at $\hat r=1$,
i.e., reach contact, before encountering
a Roche limit. We do not know if this has any physical significance.
Clearly, equilibrium sequences for real systems with $p\ne1$ always
terminate at the onset of Roche lobe overflow and mass transfer. This
could most naturally be associated with the Roche limit as
we define it here. It may be that our contact equilibrium solutions with
$p\ne1$
correspond to true asymmetric contact configurations that can be reached,
e.g., after a brief episode of mass transfer in a semidetached system.
However, confirmation of this must await more detailed numerical hydrodynamics
calculations including the treatment of mass transfer.

\vskip 0.3truecm
\centerline{\bf 3.3 General Classification of the Equilibrium Sequences}
\vskip 0.2truecm

Depending on the masses, radii, and polytropic indices of
the two components, different behaviors are possible for the equilibrium
sequences near contact.
We summarize all possible types of behaviors
for a synchronized system in Figure~3, where we show schematically
the equilibrium energy curves $E_{eq}(r)$, along which we locate
the points of secular instability, dynamical instability,
and the Roche limit. As discussed above, along such synchronized sequences,
the energy minimum corresponds to the onset of secular instability, while
the point of minimum $r$ corresponds to the Roche limit.

Six different possible behaviors can be distinguished:

(a) The binary encounters no stability or Roche limit prior to
contact, hence stable equilibrium solutions exist all the way to contact.

(b) The binary encounters a secular stability limit prior to contact, but no
dynamical stability or Roche limit. Beyond the secular stability limit,
the system becomes unstable on the viscous dissipation time scale, but all
equilibrium solutions are dynamically stable.

(c) The binary encounters a secular stability limit and a Roche limit before
contact, but not a dynamical stability limit. The binary at the Roche
limit is secularly unstable but dynamically stable.

(d) The binary encounters a secular stability limit and a
dynamical stability limit prior to contact, but not a Roche limit.
The system becomes secularly unstable first, then dynamically unstable.

(e) The binary encounters a secular stability limit, a dynamical
stability limit and a Roche limit prior to contact. The binary becomes
dynamically unstable before the Roche limit is reached, and thus a binary
at the Roche limit is dynamically unstable.

(f) The binary encounters a secular stability limit, a Roche limit
and a dynamical stability limit prior to contact.
The Roche limit is reached prior to the dynamical stability limit.
The binary at the Roche limit is secularly unstable but dynamically stable.

The situation for equilibrium sequences with constant circulation
is simpler. In this case, the minimum of $E_{eq}(r)$
corresponds to the dynamical stability limit. The different
behaviors can then be summarized as (a)-(c) in Fig.~3.
For~(a), equilibrium solutions all the way to contact are
dynamically stable; for (b) a dynamical instability is encountered prior
to contact; for (c) the binary first encounters a dynamical stability limit
and then a Roche limit prior to contact.

\vskip 0.5truecm
\centerline{\bf 4. APPLICATIONS}
\vskip 0.3truecm

In this section, we show how the general Darwin-Riemann configurations for
polytropes can be used to construct simple models for a variety of
different astrophysical systems. We exploit the freedom to assign different
polytropic indices $n$ and $n'$ and different constants $K$ and $K'$ for
the two components. The choice of $n$ and $n'$ allows us to model the
distribution of mass inside each component independently,
while the choice of $K$ and $K'$ allows us to mimic different types
of realistic mass-radius relations (cf.\ eq.~[2.39]).

\vfill\eject
\centerline{\bf 4.1  Models with $K=K'\,$: Low-mass White Dwarfs and Planets}
\vskip 0.2truecm

Low-mass white dwarfs with $10^{-3}\ll M/M_{\odot}\ll 1$
are supported by degenerate pressure from nonrelativistic ideal
electrons (see, e.g., Shapiro \& Teukolsky 1983). In this mass range,
the equation of state is that of a polytrope with $n=1.5$ and
$K=1.0036\times 10^{13}/\mu_e^{5/3}$ (cgs), where
$\mu_e=\langle A/Z \rangle$ is the mean molecular weight per electron,
$A$ is the atomic weight and $Z$ is the atomic number;
both $n$ and $K$ are independent of mass.
Thus for a binary consisting of two white dwarfs with masses
$M$ and $M'$ in this range, the ratio of the stellar
radii is $R_o/R_o'=(M/M')^{(1-n)/(3-n)}=(M/M')^{-1/3}$.

Models with $K=K'$ can also be applied crudely to cold
objects with sufficiently low mass, $M\ll 10^{-3}M_{\odot}$
(Zapolsky \& Salpeter 1969; Lai, Abrahams \& Shapiro 1991), such
as planets and their satellites.
For these objects, solid state (Coulomb) forces render the
equation of state quite stiff and the
density nearly uniform, i.e., the configurations
are nearly incompressible with $n\simeq 0$. Thus for binary
systems containing these objects, we can take $R_o/R_o'=(M/M')^{1/3}$.

Some equilibrium properties for systems with $K=K'$ were shown
in Figure~1. The systems are assumed to be synchronized ($f_R=f_R'=0$), and
have $n=n'=1.5$. Different curves correspond to
different mass ratios: $p=1$, 0.8, 0.6, and 0.5. All sequences terminate
at contact, where $\hat r=1$. We see that for $n=1.5$,
all sequences encounter a secular stability limit (minimum of $E$ and $J$)
prior to contact. When $M$ and $M'$ are sufficiently different,
a Roche limit is found prior to contact.

In Table~1 we present selected equilibrium sequences for synchronized
Darwin-Riemann binaries with $K=K'$ and $n=n'$.
Each sequence is parametrized by monotonically
decreasing values of the parameter $\hat r$ (cf.\ eq.~[2.50]).
All sequences terminate at contact ($\hat r=1$) or at the Roche limit.
For each equilibrium solution, we list various physical properties of interest,
including the ratios $a_2/a_1$, $a_3/a_1$, $R/R_o$,
and the quantities
$$\bar r={r\over R_o+R_o'},~~~~
\bar \Omega={\Omega\over (\pi G\bar \rho_o)^{1/2}},~~~~
\bar J={J\over (GM^3R_o)^{1/2}},~~~~
\bar E={E\over (GM^2/R_o)},  \eqno(4.1)$$
where ${\bar\rho}_o\equiv M/(4\pi R_o^3/3)$.

The three critical points, i.e., the
secular stability limit $r_{sec}$, the dynamical stability limit $r_{dyn}$
and the Roche limit $r_{lim}$, along an equilibrium sequence
can be calculated using the method discussed in
\S 3. In Table~2 we give some results for the synchronized configurations
($f_R=f_R'=0$) with $n=n'=0,~0.5,~1.5,~2.5$.
Several different values of $p=M/M'$ are considered for each $n$.
At each of these critical points, various physical quantities are given.
Note that for given $n$ and $p$, not all of these critical points exist
prior to contact.
When they do exist, we list them in the order of decreasing $\hat r$.
As noted before, along an equilibrium sequence,
$\hat r$ monotonically decreases, whereas $\bar r$ can increase
beyond the Roche limit. Therefore the critical point
with larger $\hat r$ occurs prior to that with smaller $\hat r$.
The existence of these critical points and the order of their appearance
(cf.\ Fig.~3) depend on the values of $n$ and $p$.

Consider the two interesting cases with $n=0$ and $n=1.5$.
For the n=0 case, both the secular and dynamical stability limits
always exist. When $p<0.79$, a Roche limit also appears.
The dynamical stability limit appears prior to the Roche limit,
except when $p<0.0042$, for which a
Roche-Riemann model is well adequate to describe the binary (see below).
Now consider the $n=1.5$ case. For $p=1$, only a secular
stability limit exists prior to contact. As the mass ratio decreases
to $p<0.76$, a dynamical instability arises; when
$p<0.745$, a Roche limit also appears. For $0.267<p<0.745$,
the dynamical instability is encountered prior to the Roche limit;
but for $p<0.267$, the Roche limit is encountered first.

To summarize, we show in Figure~4 (i) a $p-n$ diagram, distinguishing
the different behaviors illustrated in Figure~3. The figure
treats synchronized configurations only. The diagram was constructed as
follows.
For given $n$ and $p$, we first determine whether a critical point
exists prior to contact.
For this we only need to solve for two neighboring equilibrium
configurations around $\hat r=1$, and compare the values of $\bar r$, $E_{eq}$
(or $J_{eq}$) for these two neighboring solutions (cf.~\S 3 and Fig.~3).
For example, if $\Delta\bar r/\Delta\hat r>0$ at $\hat r=1$, then a Roche
limit does not occur prior to contact; otherwise a Roche limit exists.
When the dynamical stability limit and the Roche limit both exist,
we need to determine which one occurs first by comparing the values of
$\hat r$ at which they occur. In Figure~4 (i), only the boundary line between
regions (e) and (f) requires solving for the values of
$\hat r_{dyn}$ and $\hat r_{lim}$.
{}From Figure~4 (i), we see that for the binaries to be dynamically stable
at the Roche limit, the stars must be sufficiently compressible (large $n$),
or the masses of the two components must be significantly different. Moreover,
as $n\rightarrow 3$, a Roche limit always occurs before contact.

A similar diagram for irrotational configurations with $K=K'$
 is shown in Figure~5 (i).
Here the Roche limit configuration, if exists at all, is always dynamically
unstable.

When $K=K'$, the less massive companion always suffers a larger
tidal deformation. This can be seen from Table~1 by comparing
the axis ratios of the two components
(e.g., $a_2/a_1$ vs.\ $a_2'/a_1'$). When $p=M/M'$ is sufficiently
small, the more massive star suffers little tidal deformation and can be
treated as a point mass. The binary can then be modeled as a
Roche-Riemann configuration (see \S 4.4).

\vskip 0.3truecm
\centerline{\bf 4.2 Models with $R_o/R'_o=M/M'\,$: Low-mass Main-Sequence
Stars}
\vskip 0.2truecm

Low-mass main-sequence (MS) stars with $0.1M_{\odot}\lo M\lo 0.8 M_{\odot}$
have extensive convective envelopes and can be modeled approximately
as polytropes with $n\simeq1.5$--3. For reference,
we have constructed polytropic models
of low-mass, Pop II MS stars by simply matching the radius and ratio
of central to mean density obtained in detailed stellar structure
calculations (D'Antona 1987).
In Table~3 we list the radii and effective $n$ obtained for
different masses. The mass-radius relation can be fitted approximately
by a power-law relation $R_o\propto M^{\alpha}$, with $\alpha \simeq 0.8$--1.
Here, for simplicity, we adopt $\alpha=1$ and we model the stars as
polytropes with $n=n'$, and with $K$ and $K'$ adjusted
to obtain $R_o/R_o'=M/M'$. We assume synchronization, since the effective
viscosity in convective envelopes is very large (Zahn 1977).

In Figure~4 (ii), we show the $p-n$ diagram distinguishing
the different types of equilibrium sequences for this model. In particular
for $n=1.5$, we find that secular instability always sets in prior to contact
or the Roche limit. When $p<0.56$, dynamical instability can also arise.
These results are only partially confirmed by more accurate, fully numerical
calculations, which indicate that Roche lobe overflow occurs before
any instability when the mass ratio $p$ is sufficiently close to
unity (see \S 5). Dynamical instabilities in models with $n=n'=1.5$ have
been identified in fully numerical
solutions, but only for equilibrium configurations where
the two stars are in deep contact
(see Rasio 1993; Rasio \& Shapiro 1993).

\vskip 0.3truecm
\centerline{\bf 4.3 Models with $R_o=R'_o\,$: Neutron Stars and Brown Dwarfs}
\vskip 0.2truecm

The internal structure of neutron stars is
determined by the nuclear equation of state (EOS)
(see, e.g., Shapiro \& Teukolsky 1983).
They are generally characterized by a very stiff
polytropic index $n \simeq 0.5$--1 for masses well above
the minimum mass ($M\go 0.1M_{\odot}$).
For many realistic EOS, the stellar radius is not very sensitive
to the mass for an appreciable mass range. For example, the recent nuclear EOS
of Wiringa, Fiks \& Fabrocini (1988)
gives a value of $R_o\simeq 10\,$km almost independent of the mass for
$0.8 M_{\odot} \lo M\lo 1.5 M_{\odot}$. Thus for a simple description of binary
neutron stars we can adopt $R_o/R_o'=1$ and $n=n'\simeq 0.5$--1.

For brown dwarfs with $0.001\lo M/M_{\odot}\lo 0.08$, the dependence
of radius on mass is also very weak (Zapolsky \& Salpeter 1969;
Lai et al.\ 1991; Burrows \& Liebert 1993).
Therefore models with $R_o/R'_o=1$ can also be used to describe
approximately brown dwarf binaries, although here $n\lo 1.5$.

Figure~4 (iii) shows the $p-n$ diagram for synchronized
configurations with $R_o/R_o'=1$ and $n=n'$.
The similarity with Figure~4 (i) is expected, since here also the more massive
star suffers less tidal deformation (see \S 4.5).
Figure~5 (ii) shows the analogous diagram for irrotational configurations.

The evolution of most close binary systems that are observed may be tracked
along
sequences of synchronized equilibrium configurations.
But this may not be true for neutron star binaries undergoing
orbital decay due to gravitational radiation emission
Bildsten \& Cutler 1992; Kochanek 1992). In this case, the viscous dissipation
time
 may never be small enough to achieve synchronization.
Therefore binary neutron stars may actually evolve
along irrotational equilibrium sequences (Kochanek 1992). We see from
 Figure~5 (ii) that all
irrotational sequences contain a dynamical stability limit as long as
$n\lo 1.2$. Thus all neutrons star binaries (except those containing
very low-mass components) encounter a dynamical instability during their
evolution.
In particular, for typical $n=0.5$--1, we see that the Roche limit
configuration is always dynamically unstable. On the basis of this result
(confirmed by recent fully numerical calculations; see Rasio \& Shapiro 1993),
we argued in LRS3 that stable mass transfer in binary neutron stars
(as proposed by Clark \& Eardley 1977;
see also Jaranowski \& Krolak 1992) cannot occur.
The only exception would be when the mass of one of the two components is
very small ($M\lo 0.4M_{\odot}$) and the viscosity is sufficient to maintain
synchronization. Indeed, we see in Figure~4 (iii) that
dynamically stable Roche limit configurations can exist for synchronized
systems only when $n\go 1.5$, or when the mass ratio $p$ is very small.

The irrotational sequences correspond
to neutron stars with no intrinsic spin (see eq.~[2.54]). When the stars
have nonzero spins, the appropriate sequences are those with constant
$\cc$ and $\cc'$ (\S2.3.3). Examples of such sequences are tabulated in LRS3,
where a more realistic model for neutron stars has also been considered.
In this more realistic model, the radius and effective polytropic index
 are obtained from the nuclear EOS of Wiringa et al.\ (1988).

\vskip 0.3truecm
\centerline{\bf 4.4 Extreme Mass Ratios: Roche-Riemann Binaries}
\vskip 0.2truecm

Let us compare the tidal deformation of the two components
in a general Darwin-Riemann configuration.
The ellipticity of $M$ due to the tidal field of $M'$ can be estimated as
$$\eps \sim {\Delta R \over R}\sim
 {M'\over M}\biggl({R_o\over r}\biggr)^3.\eqno(4.2)$$
Similarly for $M'$,
$$\eps'\sim {M\over M'}\biggl({R'_o\over r}\biggr)^3.\eqno(4.3)$$
Clearly we have
$$\eps'\sim\eps\biggl({M\over M'}\biggr)^2\biggl({R_o'\over R_o}\biggr)^3
=\eps \biggl({M\over M'}\biggr)^{2-3\alpha}
=\eps p^{2-3\alpha},\eqno(4.4)$$
where we have assumed that $R_o/R_o'=(M/M')^{\alpha}$.
We see that when $\alpha<2/3$, $\eps'\rightarrow 0$ as
the mass ratio $p=M/M'\rightarrow 0$,
i.e., the more massive star suffers no tidal deformation
and behaves like a point mass in this limit. Therefore
when $\alpha<2/3$, our Darwin-Riemann solutions should approach
the Roche-Riemann solutions (LRS1, \S8) for sufficiently small $p$
(and as long as spin effects can be neglected).
This is confirmed by our numerical calculations.

Consider first the models with $K=K'$ and $n=n'$ (\S 4.1).
For these models we have $\alpha=(1-n)/(3-n)$,
so the inequality $\alpha<2/3$ is always satisfied. Thus for small $p$,
these models become Roche-Riemann configuration.
For example, in the model of a planet-satellite system, we can take
$n=0$ so that $\alpha=1/3$, and we have $M_{planet}\gg M_{satellite}$,
so that the tidal deformation of the planet is negligible.
The planet can therefore be represented by a point mass, even though
 its size is much larger than that of the satellite.
The binary neutron star models with $R_o=R_o'$ discussed in \S 4.3 also
have this asymptotic property.

In contrast, for typical low-mass MS stars with $0.8\lo\alpha\lo 1$
(\S 4.2), we have the opposite result: the more massive star
suffers a larger tidal
deformation than the less massive star as a result of its much larger size.
In fact, in this case, as $p\rightarrow 0$, we approach the extreme
opposite limit of a Roche-Riemann configuration, i.e.,
a test particle around a massive companion (see LRS1, \S8.3).

As discussed in LRS1, irrotational Roche-Riemann configurations
always encounter a dynamical stability limit (minimum of $E$ and $J$) followed
by a Roche limit (minimum of $r$), while
the synchronized Roche configurations always encounter a secular
stability limit followed by a Roche limit. The existence of a
dynamical instability for Roche configurations depends on the
values of $p$ and $n$.
We show in Figure~6 the boundary line between the two regimes.
As expected, this line coincides exactly with the long-dashed
lines in Figure~4 (i) and~(iii) in the limit of small $p$.
Note that for $p=0$ (e.g., a star orbiting synchronously
around a supermassive black hole), the Roche limit is
always encountered prior to the dynamical stability limit.

\vskip 0.5truecm
\centerline{\bf 5. COMPARISON WITH RECENT NUMERICAL WORK}
\vskip 0.3truecm

To assess the validity of our binary equilibrium models,
we have performed extensive
comparisons with the results of fully numerical studies
by Hachisu and Eriguchi (1984b, herafter HE), who used a
grid-based technique in three dimensions, as well as our own
recent calculations using the smoothed particle hydrodynamics
(SPH) method (see Monaghan 1993 for a recent review).
The SPH method is used to construct hydrostatic equilibrium
configurations by letting a system containing initially two
spherical polytropes relax in the presence of artificial friction
forces. The binary separation $r$ is maintained constant during
the calculation, which is performed in the corotating frame of the binary.
See Rasio \& Shapiro (1992, 1993) for details.
Similar comparisons were presented in LRS1 for binary models containing
two identical components. Here we present selected results for
binary models with mass ratio $p\ne1$. The comparison is limited to
synchronized models (Darwin configurations) since no numerical
data are available for nonsynchronized configurations in three
dimensions. While the self-consistent-field method of HE
can only give hydrostatic equilibrium solutions, the SPH method
can be used to test directly the dynamical stability of the solutions.
This is done simply by using the equilibrium configuration as an
initial condition for a dynamical calculation. Unstable systems
evolve to a rapid coalescence and merging of the two components
in just a few orbital periods, while stable systems maintain their
circular orbit indefinitely (Rasio \& Shapiro 1992, 1993).

In Figure~7 we compare our ellipsoidal results for incompressible
Darwin models ($n=n'=0$, $f_R=f'_R=0$) with $p=1.5$, 5, and~10, to those of HE.
Following HE here for convenience, we show the variation of $\Omega^2$ as a
function
of $J$, and we adopt the units defined in
equations~(4.1) (note that $R=R_o$ in the incompressible case).
We note immediately that
{\it both calculations agree in predicting the existence of a minimum of
$J$ prior to the Roche limit\/}. This is an important point, which gives
us complete confidence that the secular instability identified
here is real and not a mathematical artefact caused by our simplifying
approximations.
Indeed, HE interpret correctly
the existence of the minimum in terms of a secular instability,
which they refer to as the ``gravogyro'' instability.
Quantitatively, the agreement between our results and those of
HE is excellent away from the Roche limit.
Very near the Roche limit, the truncation of the interaction potential
to quadrupole order introduces nonnegligible errors in our models.
As a result, the Roche limit appears slightly earlier along the numerically
determined sequence. As expected, the deviations are smaller for systems with
more extreme mass ratios where the Roche limit is at larger $r$.
Note that the Roche limit also appears as a turning point (where
$d\Omega^2/dJ=0$)
followed by a (very short) second branch of equilibrium solutions
in the calculations of HE. As discussed in \S3.2, we believe this to
be the result of truncating the interaction potential to
some high but finite order in a multipole expansion.
Unfortunately, the numerical calculations of HE were limited to
the incompressible case (Hachisu \& Eriguchi 1984a and Hachisu 1986b
have considered compressible binary models, but only with $p=1$).
Rasio \& Shapiro (1993) have recently performed a series of calculations
for compressible systems with $p\ne1$ using the SPH method. As far as we are
aware, these are the first three-dimensional calculations for
compressible binary systems with $M\ne M'$.

In Figure~8 we present a comparison with these recent SPH results
for models of main-sequence star binaries with
$n=n'=1.5$ and $R_o/R_o'=M/M'$ (cf. \S 4.2). For $p=1$ it is possible to
extend the equilibrium sequence calculated with SPH all the way
into deep contact. This sequence (Fig.~8a) terminates when mass shedding occurs
through the outer Lagrangian points. In contrast, all sequences with
$p=M/M'\ne1$ terminate at the onset of Roche lobe overflow, beyond which
an equilibrium solution no longer exists for that value of $p$.
In all cases, the equilibrium $J(r)$ curves determined by the two
methods are in excellent quantitative agreement all the way
to the onset of Roche lobe overflow or point of first contact
(for $p=1$), as determined by SPH. An important conclusion is that,
{\it for sufficiently compressible fluids\/}, the much simpler quasi-analytic
results derived here can be used
even for applications requiring high quantitative accuracy,
as long as the precise location of the Roche limit is not required.
A similar situation was encountered in LRS1 (cf. \S 3 of that paper)
 with respect to the mass
shedding limit along equilibrium sequences of single rotating stars.
The energy variational method can be used to determine equilibrium
properties of uniformly rotating polytropes quite accurately up to the mass
shedding limit, but it cannot by itself predict accurately
the position of the mass shedding limit.
Note also in Figure~8 that the numerically determined terminal
configurations, corresponding to the onset of Roche lobe overflow
and mass transfer, are secularly unstable (past the minimum of $J$) only for
sufficiently small mass ratios ($p\lo0.5$). This is in contrast with the
prediction of \S4.2, that all Roche limit configurations should be secularly
unstable in this case. In addition, all SPH solutions with $0.25\le p<1$
remain dynamically stable all the way to the Roche limit.
The sequence of contact solutions with $p=1$ (beyond the point marked C in
Fig.~8a) encounters a dynamical instability prior to the mass shedding
limit (onset of mass loss through the outer Lagrangian points; see
Rasio \& Shapiro 1993 for details).

In Figure~9, we show a similar comparison with SPH for models
of a binary containing two highly-incompressible degenerate stars
with $n=n'=0.5$, $K=K'$, and $p=0.85$.
{\it The two methods agree in predicting that the Roche limit configuration is
both secularly and dynamically unstable\/}.
Quantitatively, the agreement is excellent prior to the minimum of $J$
(deviations
are $\ll1$\% between the two $J(r)$ curves), and slightly less good
($\sim1$\%) from the minimum of $J$ to the Roche limit.
The values of $r_{sec}/R_o$ predicted by the two methods agree to
within 5\%, the values of $r_{lim}/R_o$ within 10\%.
We conclude that, {\it for sufficiently incompressible fluids\/},
the quantitative
accuracy of the ellipsoidal solutions is always excellent, both for
determining the global physical properties along an equilibrium sequence
and for locating the onset of instability and Roche lobe overflow.

\vskip 0.5truecm
\centerline{\bf 6. BINARY EVOLUTION TRACKS}
\vskip 0.3truecm

In this section, we show how the equilibrium sequences calculated
in \S 4 can be used to describe the secular orbital evolution of
close binary systems in the presence of dissipation.
We consider two astrophysically important dissipation
mechanisms that can drive orbital evolution: viscosity and gravitational
radiation. We first derive the secular rates of dissipation due to these
mechanisms for binaries in circular orbit. Our treatment is exact in the
ellipsoidal approximation.

\vfill\eject
\centerline{\bf 6.1 Dissipation Rates and Time Scales}
\vskip 0.2truecm

\bigskip
\centerline{\it 6.1.1 Viscous Dissipation}
\medskip

In any fluid system, the rate of energy loss due to shear
viscosity is given by (Landau \& Lifshitz 1987)
$$\dot E_v=-\int\!\sigma_{ij}v_{i,j}\, d^3x,\eqno(6.1)$$
where $v_i$ is the fluid velocity and $\sigma_{ij}$
is the viscous stress tensor,
$$\sigma_{ij}=\eta\, \left( v_{i,j}+v_{j,i}-{2\over3}\delta_{ij}
   v_{k,k}\right).\eqno(6.2)$$
We denote by $\eta=\rho\nu$ the dynamical shear viscosity, where $\nu$ is the
kinematic shear viscosity (units cm$^2/$s).
Consider the viscous dissipation in the star of mass $M$.
Using equation~(2.14) with ${\bf u^{(0)}}={\bf v}$ here, we get
$$\eqalign{
v_{1,2} &={a_1\over a_2}\Lambda-\Omega,\cr
v_{2,1} &=-{a_2\over a_1}\Lambda+\Omega,\cr
v_{i,j} &=0~~~~~{\rm otherwise}.
}\eqno(6.3)$$
Thus the rate of viscous energy dissipation is given by
$$\dot E_v(M)=-{\bar \nu}M\Lambda^2\left({a_1^2-a_2^2\over a_1a_2}\right)^2,
\eqno(6.4)$$
where $\bar\nu={1\over M}\int\!\nu\, dm$ is the mass-averaged viscosity.
The viscous dissipation rate $\dot E_v(M')$ in the star of mass $M'$
can be written down similarly.
We see that $\dot E_v(M) \propto \Lambda^2$, so that
the viscous dissipation vanishes for synchronized configurations, as expected.

Since viscous forces conserve the total angular momentum, we have $\dot J_v=0.$
{}From equation~(2.52), we then
obtain the secular rate of change of circulation in $M$,
$$\dot\cc_v(M)={\dot E_v(M)\over \Lambda}
=-{\bar \nu}M\Lambda \left({a_1^2-a_2^2\over a_1a_2}\right)^2,
\eqno(6.5)$$
and a similar expression for $M'$.

\bigskip
\centerline{\it 6.1.2 Gravitational Radiation}
\medskip

In the weak-field, slow-motion limit of general relativity, the total
rate of energy loss due to
gravitational wave emission is given by (cf. LRS3)
$$\dot E_{GW}=-{32G\over 5c^5}\,\Omega^6\,(\mu r^2)^2\,
\biggl [1+{1\over \mu r^2} (I_{11}+I_{11}'-I_{22}-I_{22}')\biggr]^2.
\eqno(6.6)$$
This is essentially the quadrupole formula (see, e.g., Shapiro \&
Teukolsky, Chap.~16) written for a binary system containing two ellipsoids.
Fluid circulation is conserved by the gravitational radiation reaction
forces, i.e., $\dot\cc_{GW}=\dot\cc'_{GW}=0$ (Miller 1974).
{}From equation~(2.52), the secular rate of angular momentum loss is given by
$$\dot J_{GW}={\dot E_{GW}\over\Omega}.\eqno(6.7)$$

Other energy and angular momentum loss mechanisms such as magnetic braking
and mass loss from the system are far more difficult to model
and will not be considered here. However, we expect
our qualitative discussion below to hold quite generally.

\bigskip
\centerline{\it 6.1.3 Time Scales}
\medskip

There are four physically distinct time scales relevant to
the evolution of the binaries: the internal hydrodynamical
time, $t_{dyn}\equiv (R_o^3/GM)^{1/2}$; the orbital period,
$P=2\pi/\Omega$; the time scale for the fluid circulation to change,
$t_{\cc}\equiv |\cc/\dot\cc|$; the time scale for angular
momentum loss, $t_J\equiv |J/\dot J|$. Most binary systems
have $t_{dyn}<P<t_{\cc}<t_J$.

For angular momentum loss due to gravitational wave emission,
$t_J$ is given by (cf.\ eqs.~[6.6] and~[6.7])
$$t_J = {5c^5\over 64G^3}{r^4\over MM'(M+M')}.\eqno(6.8)$$
To estimate $t_{\cc}$, we use equations~(2.20) and~(2.54) to write
$\Lambda \simeq \Omega -\Omega_s$ for large $r$.
The angular frequency $\Lambda$ thus
measures the departure from synchronization. From equation~(6.5) we then
get
$$t_{\cc} \simeq \left({a_1a_2\over a_1^2-a_2^2}\right)^2
\left({R_o^2\over\bar\nu}
\right){\Omega_s\over\Lambda}.\eqno(6.9)$$
Note that $t_{\cc}$ is propotional to the viscous dissipation time $t_{vis}
\equiv R_o^2/\bar\nu$. Note also that exact synchronization
can never be achieved, since $t_{\cc}\rightarrow \infty$
as $\Lambda\rightarrow 0$.

Using $a_1\sim a_2\sim R_o$ and
$(a_1-a_2)/R_o \sim (M'/M)(R_o/r)^3$, equation~(6.9) gives
$$t_{\cc} \sim {\Omega_s\over \Lambda} t_{syn}
\eqno(6.10)$$
where we have defined the synchronization time
$$t_{syn}\equiv \left({M\over M'}\right)^2\left({r\over R_o}\right)^6
{(GMR_o)^{1/2}\over \bar\nu}\left({R_o^3\over GM}\right)^{1/2}.
\eqno(6.11)$$
This scaling for $t_{syn}$ agrees with the results of
the standard weak-friction model of tidal interactions
(Alexander 1973; Zahn 1977). For solar-type MS stars,
the average plasma viscosity is of order $10^3\,$cm$^2/$s,
corresponding to $\bar\nu/(GMR_o)^{1/2}
\sim 10^{-15}$. This is far too small to explain the high degree of
synchronization observed in close MS star binaries.
However, turbulent viscosity associated  with convection in late-type systems
can give $1\go \bar\nu/(GMR_o)^{1/2}\gg 10^{-15}$ (Zahn 1977).
Possible sources of anomalous viscosity in degenerate stars are
discussed by Kochanek (1992).

\vskip 0.3truecm
\centerline{\bf 6.2 Orbital Evolution}
\vskip 0.2truecm

\bigskip
\centerline{\it 6.2.1 Basic Assumptions}
\medskip

Depending on which mechanism dominates, the time scale for
energy loss according to equation~(2.52)
can be either $t_J$ or $t_{\cc}$, i.e., $t_E\equiv
|E/\dot E|\sim {\rm min}(t_{\cc},t_J)$.
We only consider regimes where both $t_J$ and $t_{\cc}$ are much longer
than $P$ and $t_{dyn}$, so that the binary evolves quasi-statically
along an equilibrium sequence. This applies to the majority of observed binary
systems, with the possible exception of neutron star binaries near coalescence.
The secular rate of change of the binary separation can then
be written
$$\dot r={{\dot E}\over {dE_{eq}/dr}},\eqno(6.12)$$
and the orbital evolution time $t_r$ is given by
$$t_r \equiv {r\over |\dot r|}={r|dE_{eq}/dr|\over |\dot E|}.\eqno(6.13)$$
Normally, when $|dE_{eq}/dr| \sim |E/r|$, we have $t_r\sim t_E$, i.e., the
orbit decays on the energy loss time scale.
But it is important to note that $t_r$ can become significantly smaller
than $t_E$ as $dE_{eq}/dr \rightarrow 0$. The quasi-static description of the
orbital
evolution becomes invalid when a dynamical stability is approached.

There are two regimes of interest.
When the fluid viscosity is sufficiently small so that
$t_{syn}\gg t_J$, the evolution of a binary system
as it loses angular momentum proceeds along constant-circulation
equilibrium curves such as those shown in Figure~3.
The values of $\cc$ and $\cc'$ are determined by
the initial spins of the stars (cf.\ eqs.~[2.54]).
The orbit decays on the time scale
$t_E\sim t_J$ until a dynamical instability is encountered, followed
by rapid coalescence and merging of the two stars (Rasio \& Shapiro 1992,
1993).
This regime is most relevant to the coalescence of binary neutron
stars driven by gravitational radiation. Indeed, for these systems,
 it has been argued that the viscosity is always too small to maintain
synchronization (Bildsten \& Cutler 1992; Kochanek 1992).
We refer the reader to our discussion of this
regime in LRS3 for more details.

Here we focus on the opposite regime when viscosity dominates and
$t_{syn}\ll t_J$. This regime applies to most observed binary
systems containing at least one nondegenerate component.
The inequality can be written
$${t_{syn}\over t_J} \sim \left({r\over R_o}\right)^2
\left({GM\over c^2R_o}\right)^{5/2}{(GMR_o)^{1/2}\over\bar\nu}\ll 1.
\eqno(6.14)$$
In this regime, binaries evolve along constant$-J$ equilibrium
sequences. In general, for a fixed value of $J$, the equilibrium
sequence is not unique: even at large separation, fixing $J$ and $r$
determines only the sum of the two spin angular momenta.
However, if a complete initial configuration
is specified with given values of $\cc$ and $\cc'$ (or $f_R$
and $f_R'$), then the constant$-J$ sequence passing through
the initial configuration can be constructed uniquely provided we know
the ratio of the averaged viscosities within the two components.

\bigskip
\centerline{\it 6.2.2 Evolution along Constant-$J$ Sequences}
\medskip

For simplicity, we treat only the case where the binary contains two
identical components. In Figure~10 we show examples of constant$-J$
equilibrium curves for $n=1.5$ and $p=1$. Three values of $J$ are
considered. We also show the synchronized (Darwin)
sequence for comparison. We see
that there exists a critical value $J=J_{min}$,
equal to the minimum of $J$ along the synchronized sequence,
above which a constant$-J$ sequence intersects the synchronized sequence.
Moreover, this intersection point exactly coincides with an
energy extremum along the constant$-J$ sequence.
This is easy to understand: from equation~(2.52) we have
$dE=2\Lambda\, d\cc$ along a constant$-J$ sequence; thus when $\Lambda=0$
(at synchronization), $dE=0$.
For the case shown in Figure~10, the intersection point (point~B
in Fig.~10a)
lies on the secularly stable branch of the synchronized sequence.
Therefore, it is the point of
minimum energy along the corresponding constant$-J$ sequence,
i.e., among all those configurations with the same $J$, the synchronized
configuration is the one that has the lowest energy.

Note that if we consider a value of $J$ just slightly greater than $J_{min}$,
the constant$-J$ curve can intersect the
synchronized curve twice, once on the secularly stable branch and
once on the unstable branch (Fig.~10b). Both intersection points
correspond to a local energy extremum along the constant-$J$ sequence.
The intersection with the stable branch is a minimum (point~H in Fig.~10b),
whereas the intersection with the unstable branch is a maximum (point~I in
Fig.~10b). This explains very clearly the physical nature of the secular
instability. Viscosity will drive a secularly unstable equilibrium
configuration {\it away\/} from synchronization at first. As a result,
the orbit can either expand (along IH) as
the system is driven towards a lower-energy, stable synchronized state,
or it can decay (along IJ) as the stars are driven to coalescence.

For the limiting case where $J=J_{min}$, the constant$-J$
sequence passes through the secular stability limit point along the
synchronized sequence (point~C in Fig.~10a).
In this case, the intersection is a stationary
point ($dE_{eq}/dr=0$, $d^2E_{eq}/dr^2=0$) of the $E_{eq}(r)$ curve
for the the constant$-J$ sequence. Note that the first secularly
unstable synchronized configuration at point~C will always be driven
to coalescence by viscous dissipation (in contrast to unstable
configurations beyond C, for which the orbit can evolve either way).
When $J<J_{min}$, the constant$-J$ sequence
never intersects the synchronized sequence, and the energy decreases
monotonically as $r$ decreases. Therefore, configurations with $J<J_{min}$
can never reach synchronization, and are always driven to coalescence
by viscous dissipation.

Clearly, the orbital evolution of a binary system depends critically on
its initial angular momentum $J_i$. When $J_i<J_{min}$, as the
the binary losses energy due to viscous dissipation, it simply slides
down the constant$-J$ curve (e.g., along FG in Fig.~10a).
The time scale for this orbital decay is the synchronization time,
since $t_r\sim t_{\cc}\sim t_{syn}$ in this case. When $J_i>J_{min}$,
the binary first evolves toward a (stable) synchronized
configuration. If the star initially spins
faster than the synchronized rate (point~E in Fig.~10a),
the orbit expands as the system evolves toward synchronization (along EB).
The Earth-Moon system is a well-known example of such an evolution.
If the initial spin is slower than the synchronized
rate (point~A), then the separation becomes smaller as
the binary evolves toward synchronization (along AB).
This evolution takes place on a time scale
$t_r\sim t_{\cc}$. Initially $t_{\cc}\sim t_{syn}$; but
$t_{\cc}$ increases as synchronization is approached (cf.\ eq.~[6.9]).
As viscous energy dissipation falls asymptotically to zero
($\dot E_v\propto \Lambda^2$), the orbit will cease to evolve if there
is no other dissipation mechanism.

\bigskip
\centerline{\it 6.2.3 Final Approach to Instability}
\medskip

Now consider what happens to a nearly synchronized system
if it is losing angular momentum by some additional process like
gravitational radiation. As the orbit decays, the binary will very
closely track the synchronized curve (along BC in Fig.~10a)
 with $\Omega\simeq \Omega_s$ and evolve toward a lower energy
and angular momentum state. This evolution takes place on
the angular momentum loss time scale, with
of $t_r\sim t_J\gg t_{syn}$. The degree of synchronization
can be estimated by setting $t_r\sim t_J\sim t_{\cc}$,
which yields $\Lambda/\Omega \sim t_{syn}/t_J \ll 1$.
Viscous dissipation is negligible during this phase,
 as $\dot E_v \sim \cc\Lambda/t_{\cc} \sim
I\Omega\Lambda/t_J \sim \dot E_{GW}(\Lambda/\Omega)$.
Equation~(6.12) applies, with $E_{eq}(r)$ calculated
for the synchronized sequence (\S 2.3.2) and $\dot E=\dot E_{GW}$ (eq.~[6.6]).

As the binary approaches
the energy minimum (i.e., the secular stability limit, at point C),
the orbital decay time $t_r$ become shorter and shorter as
$dE_{eq}/dr\rightarrow 0$ (cf. eq.~[6.13]). At some point prior to the
energy minimum, $t_r$ becomes comparable to $t_{syn}$,
i.e., the orbit decays so fast that viscosity can no
longer maintain synchronization. Clearly, the evolution cannot follow the
synchronized sequence beyond the point of energy minimum.
As $\Lambda$ becomes comparable to $\Omega$, viscous dissipation resumes and
eventually
becomes dominant, with $\dot E_v\gg \dot E_{GW}$. Note that viscosity
is now driving the system {\it away from synchronization\/}.
The subsequent and final evolution follows closely
the constant-$J$ curve with $J=J_{min}$
(along CD in Fig.~10a). Thus the final coalescence takes place on a time scale
$t_r\sim t_{\cc}\sim t_{syn}\ll t_J$. It is driven almost entirely by
internal viscous dissipation, with angular momentum loss playing a
negligible role.

The terminal phase of the orbital decay outlined above requires
modification when a dynamical stability limit exists. Consider the binary
sequences
with $p=1$ and $n=0$ shown in Figure~11. We consider again three constant-$J$
sequences with $J>J_{min}$, $J=J_{min}$ and $J<J_{min}$, where $J_{min}$
is the minimum angular momentum along the synchronized sequence.
A dynamical stability limit is encountered prior to contact for all these
sequences. Along a constant-$J$ sequence, the dynamical
stability limit corresponds to an extremum in both energy and circulation
$\cc$,
as $dE=2\Lambda\, d\cc=0$. This is analogous to the dynamical
stability limit along a constant$-\cc$ sequence, where $dE=dJ=0$.
As in Figure~10, for $J>J_{min}$,
the constant-$J$ curves also intersect the synchronized sequence
where they reach an energy extremum (as $dE=2\Lambda d\cc$ with $\Lambda=0$).
Therefore, there are now {\it three\/} extrema along a constant-$J$
sequence with $J>J_{min}$.

Now consider the orbital evolution of the systems shown in Figure~11.
When $J>J_{min}$, viscosity first drives
the binary to a synchronized state in a time $\sim t_{syn}$
(along EB or AB, depending on whether the initial spin rate is larger or
smaller than the synchronized spin rate). Then the binary follows closely the
synchronized sequence (along BC) for a time $\sim t_J$, until the secular
stability
limit is reached (point~C). Beyond that point the evolution
 follows the constant $J=J_{min}$
sequence (along CD), for a time $\sim t_{\cc}\sim t_{syn}$.
Before reaching contact, however, the system becomes
dynamically unstable (point~D), and equation~(6.12) is
no longer valid. The coalescence accelerates abruptly and the two stars
merge in a time $\sim P$. This transition to a dynamical coalescence
 has been studied in LRS2 and LRS3.
Similarly, when $J<J_{min}$, the binary slides down
the constant$-J$ curve (along HI) in a time $\sim t_{\cc}\sim t_{syn}$
and eventually becomes dynamically unstable as well.

\vskip 0.3truecm
\centerline{\bf 6.3 Comparison with Earlier Work}
\vskip 0.2truecm

The secular stability of close binary systems has been discussed previously
by Counselman (1973), Hut (1980) and others,
in the limit where both stars can be represented by rigid spheres (i.e.,
neglecting spin-induced and tidal deformations). In this limit,
the total angular momentum of a synchronized system is simply
$$J^{(S)}(r)=(\mu r^2+I_o+I_o')\left({GM_t\over r^3}\right)^{1/2},\eqno(6.15)$$
where $M_t=M+M'$, $I_o={2\over5}\kappa_nMR_o^2$ and
$I_o'={2\over5}\kappa_n'M'R_o'^2$. The corresponding total
equilibrium energy is
$$E^{(S)}(r)=-{GMM'\over 2r}+{1\over 2}(I_o+I_o'){GM_t\over r^3}
+E_{\infty},\eqno(6.16)$$
where $E_{\infty}$ is the energy at infinite separation.
The secular stability limit $r_m$ is the point at which $dE^{(S)}/dr=0$,
which gives
$$\mu r_m^2=3(I_o+I_o').\eqno(6.17)$$
The corresponding (minimum) value of $J$ is
$$J^{(S)}_{min}={4\over 3}\left[3\,G^2 \mu^3 M_t^2 (I_o+I_o')\right]^{1/4}.
\eqno(6.18)$$
Equations~(6.15), (6.17), and~(6.18) can be combined together with the
requirement that $J>J_{min}$ (cf.\ \S6.2.2) to show
that synchronization is possible only when
the orbital angular momentum exceeds the spin angular momentum
by more than a factor of three (Counselman 1973; Hut 1980).

The above analysis applies only when tidal effects are small near $r_m$,
which requires $r_m\gg R_o+R_o'$. Consider for simplicity the case
where $I_o'\ll I_o$ (e.g., when $M'$ is a point mass).
Equation~(6.17) implies that
$r_m \sim R_o (1+M/M')^{1/2}$. So the inequality $r_m\gg R_o$ requires
$M\gg M'$, i.e., the mass of the extended star must be much larger
than that of its smaller-size companion (A typical example is
provided by planet-satellite systems).
When these inequalities are not satisfied, the simple analysis
based on spheres is not valid, since tidal effects cannot be ignored.

Our general results agree with these early studies only in the limit where
$p=M/M'\gg1$. Values of $r_m$ and $J_{min}$ for
different cases can be read off Table~2, and can be
compared with equation~(6.18).
As expected, the differences are largest when $p$ is close to unity.
For example, for the case illustrated in Figure~9 (where $p=0.85$),
the two-sphere model (eqs.~[6.17] and~[6.18])
predicts $r_m/R_o\simeq2.0$ and ${\bar J}_{min}\simeq1.16$, whereas
we find $r_m/R_o\simeq3.0$ and ${\bar J}_{min}\simeq1.28$.
Even for large $p$, the rigid-sphere model can give
incorrect results at small separation.
Figure~12 shows a comparison between the two models
for a system containing an incompressible star ($n=0$) in
orbit around a point mass, with $p=M/M'=10$ ($M'$ is the point mass).
Constant$-J$ sequences with $J<J_{min}$, $J=J_{min}$ and
$J>J_{min}$ are shown. Combining equations~(6.15) and~(6.16)
and letting $I'_o=0$, we find that the total energy
along a constant$-J$ sequence is given by
$$E_J^{(S)}(r)=-{GMM'\over 2r}+{1\over 2I_o}\left[J-\mu r^2
\left({GM_t\over r^3}\right)^{1/2}\right]^2+E_{\infty}, \eqno(6.19)$$
in the rigid-sphere model.
We see in Figure~12 that this result agrees with our calculations at large
separation, but becomes invalid for small $r$. The expression correctly
predicts the existence of a minimum and maximum residing on the
synchronized curve given by equation~(6.16). However, expression~(6.19)
does not exhibit the additional (third) energy extremum which corresponds
to the onset of dynamical instability (cf.\ Fig.~11).


\vskip 1.5truecm
\centerline{\bf Acknowledgements}
\bigskip
This work has been supported in part
by NSF Grant AST 91-19475 and NASA Grant NAGW-2364 to Cornell University.
Partial support was also provided by NASA through Grant HF-1037.01-92A
awarded by the Space Telescope Science Institute which is operated by the
Association of Universities for Research in Astronomy, Inc.,
for NASA under contract NAS5-26555. Computations were performed
at the Cornell National Supercomputer Facility,
a resource of the Center for Theory and Simulation in Science
and Engineering at Cornell University, which receives major funding from
the NSF and from the IBM Corporation, with additional
support from New York State and members of its Corporate Research Institute.

\vfill\eject
\centerline{\bf APPENDIX A: ASYMPTOTIC RESULTS FOR LARGE SEPARATION}
\vskip 0.3truecm

As noted in \S 2.2, to quadrupole order, the only coupling
between the equilibrium equations for two stars is through the orbital angular
velocity $\Omega$. For sufficiently large $r$, this coupling becomes
very small (i.e., we can set $\delta\simeq0$ and $\delta'\simeq 0$ in the
equations). Thus to determine the structure of the star of mass $M$, we can
treat $M'$ as a point mass. The general Darwin-Riemann problem then reduces to
the
Roche-Riemann problem (LRS1, \S8). Here we derive asymptotic results in the
limit where
$r\gg (1+M'/M)^{1/3}R_o$ for various Roche-Riemann configurations.
References to key equations in LRS1 are  indicated with numbers preceded by an
``I''.

\def\ai {{\alpha_i}}
\def\asum {{\alpha_1+\alpha_2+\alpha_3}}
\def\Order {{\cal O}}

\bigskip
\centerline{\it Index Symbols}
\medskip

We start by obtaining expressions for the index symbols $A_i$ and
related quantities for small deviations from a spherical shape. Let's
rewrite the semi-major axes as
$$a_i\equiv R_o(1+\alpha_i),~~~~~~~{\rm with~} \alpha_i\ll1.\eqno(A1)$$
As we show below, the $\alpha_i$ are $\Order(R_o^3/r^3)$ in general
for large $r$. Expanding the integrand in the definition~(2.8) we get
$$A_i=R_o^3(1+\asum)\int_0^\infty\left[1-{R_o^2\over R_o^2+u}(\asum+2\ai)
\right]{du\over(R_o^2+u)^{5/2}} +\Order(\ai^2) \eqno(A2)$$
The integral is now elementary and we find
$$A_i={2\over3}+{4\over15}(\asum)-{4\over5}\ai +\Order(\ai^2) \eqno(A3)$$
For spheroids with $a_1=a_2>a_3$ and
$e^2\equiv1-a_3^2/a_1^2=2(\alpha_1-\alpha_3)
+\Order(\ai^2)$ this becomes
$$\eqalign{
A_1=A_2&={2\over3}-{2\over15}e^2 +\Order(e^4), \cr
    A_3&={2\over3}+{4\over15}e^2 +\Order(e^4). \cr
} \eqno(A4)$$
Related quantities of interest are (cf.\ eqs.~[2.7] and~[2.43])
$$f={1\over2}{A_1a_1^2+A_2a_2^2+A_3a_3^2\over(a_1a_2a_3)^{2/3}}
   = 1 + \Order(\ai^2), \eqno(A5)$$
and
$$\eqalign{
A_1a_1^2-A_3a_3^2&=(a_1^2-a_3^2)B_{13}={8\over15}(\alpha_1-\alpha_3)R_o^2,\cr
A_2a_2^2-A_3a_3^2&=(a_2^2-a_3^2)B_{23}={8\over15}(\alpha_2-\alpha_3)R_o^2.\cr
}  \eqno(A6)$$

\bigskip
\centerline{\it Maclaurin Spheroids}
\medskip

For future reference we first derive the $\alpha_i$ for a slowly rotating
Maclaurin spheroid (LRS1, \S3). We introduce the dimensionless angular
velocity ${\bar\Omega_s}\equiv\Omega_s/(\pi G{\bar\rho}_o)^{1/2}$ as a small
parameter in all expansions. As shown below we have
$\ai=\Order({\bar\Omega}_s^2)
=\Order(e^2)$.

We first expand expression~(I.3.21) for the ratio $T/|W|$ and find
$${T\over|W|}={2\over15}e^2 +\Order(e^4) = {4\over15}(\alpha_1-\alpha_3).
\eqno(A7)$$
Similarly, expression~(I.3.8)  gives $g=1+\Order(e^4)$.
Inserting these results into equation~(I.3.25)  gives the volume
expansion factor to $\Order(e^2)$,
$${a_1^2a_3\over R_o^3}-1=2\alpha_1+\alpha_3={8\over5}\left({n\over3-n}
\right) (\alpha_1-\alpha_3)={2\over15}\left({n\over3-n}\right) e^2. \eqno(A8)$$
This is one equilibrium condition on $\alpha_1$ and $\alpha_3$. As a
second equilibrium condition, we use the first of equations~(I.3.27), together
with expression~(I.3.28) for $\hat\Omega$. Expanding these to $\Order(e^2)$
we get
$${\hat\Omega}^2={8\over15}e^2={16\over15}(\alpha_1-\alpha_3)
=q_n{\bar\Omega}_s^2. \eqno(A9)$$
We can now solve equations~(A8) and~(A9) to obtain explicit
$\Order({\bar\Omega}_s^2)$
relations for the volume expansion factor,
$$2\alpha_1+\alpha_3={3\over2}q_n\left({n\over3-n}\right){\bar\Omega}_s^2,
\eqno(A10)$$
the eccentricity
$$e^2={15\over8}q_n{\bar\Omega}_s^2, \eqno(A11)$$
and the axes
$$\eqalign{
\alpha_1&={3\over16}q_n\left({5+n\over3-n}\right){\bar\Omega}_s^2, \cr
\alpha_3&=-{3\over8}q_n\left({5-3n\over3-n}\right){\bar\Omega}_s^2. \cr
} \eqno(A12)$$

\bigskip
\centerline{\it Roche Ellipsoids}
\medskip

Now turn to Roche ellipsoids (LRS1, \S 7).
We first use equation~(I.7.21) for the volume expansion factor,
where the term containing $\delta=\Order(R_o^5/r^5)$ can be neglected
and we can set $f=1$ (cf.~eq.~[A5]). Using
$${T_s\over|W|}={{1\over2}I\Omega^2\over\displaystyle{3\over5-n}{GM^2\over R}}
={1\over3}q_n\left({1+p\over p}\right) {R_o^3\over r^3}
+\Order\left({R_o^6\over r^6}\right)  \eqno(A13)$$
we get
$${a_1a_2a_3\over R_o^3}-1=\asum=q_n\left({2n\over3-n}\right)\,
\left({1+p\over p}\right) {R_o^3\over r^3}
+\Order\left({R_o^6\over r^6}\right).  \eqno(A14)$$
We obtain two more equations for the $\ai$ from the two equilibrium
conditions~(I.7.18) and~(I.7.19). Since
$${\tilde\mu}={GM'/r^3 \over \pi GM/({4\over3}\pi R^3)}
={4\over3}{1\over p} {R_o^3\over r^3}+\Order\left({R_o^6\over r^6}\right),
  \eqno(A15)$$
is already $\Order(R_o^3/r^3)$, the terms inside the brackets need not be
expanded. Using expressions~(A6) we then find that the two equilibrium
conditions can be written to this order as
$$\eqalign{
  \alpha_1-\alpha_3&={5\over4}q_n\left({4+p\over p}\right) {R_o^3\over r^3},\cr
  \alpha_2-\alpha_3&={5\over4}q_n\left({1+p\over p}\right) {R_o^3\over r^3}.\cr
}  \eqno(A16)$$
We now solve equations~(A14) and~(A16) for $\ai$, and find
$$\eqalign{
\alpha_1&={1\over3}q_n\left({1+p\over p}\right){R_o^3\over r^3}\left[{5\over4}
          \left({7+p\over1+p}\right)+\left({2n\over3-n}\right)\right] \cr
\alpha_2&=-{1\over3}q_n\left({1+p\over p}\right){R_o^3\over r^3}\left[{5\over4}
          \left({2-p\over1+p}\right)-\left({2n\over3-n}\right)\right] \cr
\alpha_3&=-{1\over3}q_n\left({1+p\over p}\right){R_o^3\over r^3}\left[{5\over4}
          \left({5+2p\over1+p}\right)-\left({2n\over3-n}\right)\right] \cr
} \eqno(A17)$$

It is also useful to obtain lowest-order changes in the total equilibrium
energy and angular momentum of the binary system. These are defined with
respect to a ficticious (nonequilibrium) system containing two spherical
(nonspinning) polytropes in a point-like Keplerian orbit,
$$\eqalign{
\Delta E_{eq}&\equiv E_{eq}-\left[{1\over2}\mu r^2\Omega_K^2 -{GMM'\over r}
              +\left({3-n\over3}\right)W_o\right],  \cr
\Delta J_{eq}&\equiv J_{eq} -\mu r^2\Omega_K,  \cr
} \eqno(A18)$$
where $W_o\equiv-[3/(5-n)]GM^2/R_o$ is the self-gravitational energy
of the spherical polytrope and $\Omega^2_K=G(M+M')/r^3$.
Using equation (I.7.11) for $E_{eq}$ we get
$$\eqalign{
\Delta E_{eq}
 &={1\over2}\mu r^2\Omega^2_K\delta -{GMM'\over
r}\left({2n+3\over9}\right)\delta
   +\left({3-n\over3}\right)\,W_o\left({R_o\over
r}-1\right)+\left({3-2n\over3}\right)\,T_s\cr
 &=\left[\left({3-n\over3}\right)\,W_o\left({R_o\over r}-1\right)
   +\left({3-2n\over3}\right)\,T_s\right]\left[1+\Order\left({R_o^3\over
r^3}\right)\right]. \cr
} \eqno(A19)$$
We now substitute $T_s={1\over2}I_o\Omega^2_K$, with
$I_o={2\over5}\kappa_nMR_o^2$,
 (to lowest order) and expression~(A14) for $R=(a_1a_2a_3)^{1/3}$ to find
$$\eqalign{
\Delta E_{eq}
  &=\left[{2n\over15}\kappa_nR_o^2M{G(M+M')\over r^3}
+{1\over5}\left(1-{2n\over3}\right)
    \kappa_nR_o^2M{G(M+M')\over r^3}\right] \left[1+\Order\left({R_o^3\over
r^3}\right)\right] \cr
  &={1\over2}I_o\Omega^2_K\left[1+\Order\left({R_o^3\over r^3}\right)\right].
\cr
} \eqno(A20)$$
To lowest order, this is just the naive result that one would have written down
immediately while ignoring the virial theorem.
For the angular momentum equations~(I.7.12) and~(A18) give immediately
$$\Delta J_{eq}=I_o\Omega_K\left[1+\Order\left({R_o^3\over
r^3}\right)\right].\eqno(A21)$$
It is useful to note that (1) $\Delta J_{eq}$ is considerably easier to
calculate
than $\Delta E_{eq}$, and (2) once an expression has been derived for $\Delta
J_{eq}$,
$\Delta E_{eq}$ can be obtained very simply by integrating the relation
$dE_{eq}=\Omega \,dJ_{eq}$ (cf.~LRS1, Appendix~D). Indeed, using
expression~(A21) we
find, to lowest order,
$$\Delta E_{eq}=\int\Omega_K{d\Delta J_{eq}\over dr}\,dr
               =-{3\over2}I_oG(M+M')\int{dr\over r^4}
               ={1\over2}I_o\Omega^2_K\eqno(A22),$$
in agreement with expression~(A20). For the more general configurations
considered below, the calculation of $\Delta E_{eq}$ by direct expansion
of $E_{eq}$ would be extremely tedious, and we will instead use
this shortcut.

Our results~(A17), (A20), and~(A21) are in agreement with those quoted by
Kochanek (1992, cf.\ his eqs.~[2.5] and~[2.6]), with his quantity
${\hat\Omega}_o^2$
equal to $q_n[(1+p)/p](R_o^3/r_3)$ in our notations (and his $I_{*}=I_o/2$).

\bigskip
\centerline{\it Irrotational Roche-Riemann Ellipsoids}
\medskip

{}From equation (I.8.2) with $\Lambda=2a_1a_2\Omega/(a_1^2+a_2^2)$ for
irrotational configurations we see immediately that $T_s=T_{+}+T_{-}
\propto(a_1-a_2)^2\propto R_o^2(\alpha_1-\alpha_2)^2$ and therefore that
$T_s/|W|=\Order(R_o^6/r^6)$. Thus from equation~(I.7.21) we see that there is
no change of volume to lowest order,
$$\asum=0. \eqno(A23)$$
With $Q_1=-Q_2=\Omega_K[1+\Order(R^6/r^6)]$ and $\Omega^2_K=\mu_R(1+p)$,
the two equilibrium conditions~(I.8.5) and~(I.8.6) give to this order
$$\eqalign{
\alpha_1-\alpha_3&={15\over4}q_n{1\over p}{R_o^3\over r^3},\cr
\alpha_2-\alpha_3&=0. \cr
} \eqno(A24)$$
Combining equations~(A23) and~(A24) we find
$$\eqalign{
\alpha_1&={5\over2}q_n{1\over p}{R_o^3\over r^3},\cr
\alpha_2=\alpha_3&=-{5\over4}q_n{1\over p}{R_o^3\over r^3},\cr
} \eqno(A25)$$
in agreement with the results quoted by Kochanek (1992;cf.\ his eq.~[2.7]).

Using equation~(I.8.3) with $f_R=-2$ and the definition~(A18) we get for
the change in angular momentum,
$$\Delta J_{eq}={1\over2}\mu r^2\Omega_K\delta +\left(I-4{I_{11}I_{22}\over I}
      \right)\Omega_K. \eqno(A26)$$
Since $I_{ii}=I_o(1+2\alpha_i)/2$ and
$I=I_{11}+I_{22}=I_o(1+\alpha_1+\alpha_2)$,
we see that the second term in expression~(A26) is of higher order. Using the
results~(A25) and the definition~(I.7.7) we get
$$\delta={3\over5}\kappa_n{R_o^2\over r^2}(2\alpha_1-\alpha_2-\alpha_3)
        ={9\over2}{\kappa_nq_n\over p}{R_o^5\over r^5},  \eqno(A27)$$
and equation~(A26) gives
$$\eqalign{
\Delta J_{eq}&={45\over8}{q_n\over p(1+p)}{R_o^3\ r^3}I_o\Omega_K \cr
    &={9\over4}\kappa_nq_nM'^2\left({G\over M+M'}\right)^{1/2}{R_o^5\over
r^{9/2}} \cr
} \eqno(A28)$$
We now evaluate the change in total energy using the trick described above.
Writing $\Omega=\Omega_K+\Delta\Omega$, with $\Delta\Omega=\Omega_K\delta/2$,
and expanding the relation $dE_{eq}=\Omega dJ_{eq}$ to lowest order we get
$$\Delta E_{eq}=\int\left(\Omega_K{d\Delta J_{eq}\over dr}
    +\Delta\Omega{dJ_o\over dr}\right)\,dr, \eqno(A29)$$
where $J_o=\mu r^2\Omega_K$. Combining equations~(A27)--(A29) we find
$$\eqalign{
\Delta E_{eq}&={15\over4}{q_n\over p(1+p)}{R_o^3\over r^3} I_o\Omega_K^2,\cr
   &={3\over2}\kappa_nq_nGM'^2{R_o^5\over r^6}.\cr
} \eqno(A30)$$
This does not agree with the result quoted by Kochanek (1992; see his
eq.~[2.8],
which appears to differ in both the magnitude and sign of the numerical
coefficient).

\bigskip
\centerline{\it General Roche-Riemann Ellipsoids}
\medskip

In the limit of small tidal perturbation and slow spin, the deviation of a
general Roche-Riemann ellipsoid from a sphere can be written as a linear
superposition of a pure tidal distortion and a distortion purely due to the
spin. Thus the lowest-order expressions for the principal axes are obtained
simply by adding expressions~(A12) and (A25).
The value of $\bar\Omega_s$ is determined by the condition
that the circulation ${\cal C}=-I_o\Omega_s$ at large $r$ (cf.\ eqs.~[2.54]).

The lowest-order changes in energy and angular momentum due to spin
are simply $\Delta E^{(S)}_{eq}={1\over2}I_o\Omega_s^2$
and $\Delta J^{(S)}_{eq}=I_o\Omega_s$, which are both independent of $r$.
For the calculation of gravitational radiation
phase shifts (cf, \S 4.3 and LRS3), it is
necessary to obtain the next higher-order, $r$-dependent term in the expansion
of
$E_{eq}$. This is done most easily using the same method as above, first
calculating $\Delta J_{eq}$ and then integrating $dE_{eq}=\Omega\,dJ_{eq}$.
By direct expansion of expression~(I.8.3) we find
$$\Delta J^{(S)}_{eq}=I_o\Omega_s\left[1+{3\over8}q_n\left({5+n\over3-n}\right)
   {\bar\Omega}_s^2\right] + I_o\Omega_K\left({45\over64}{q_n\over1+p}
{\bar\Omega}_s^2\right).
\eqno(A31)$$
Using equation~(A29) we then get, after some algebra,
$$\Delta E^{(S)}_{eq}={3\over32}\kappa_nq_nGMM'{\bar\Omega}_s^2{R_o^3\over r^3}
     +{\rm ~const}. \eqno(A32)$$
The total $\Delta E_{eq}$ is obtained by adding expressions~(A30)
and~(A32).

\vfill\eject
\centerline{\bf APPENDIX B: SECOND DERIVATIVES OF THE ENERGY FUNCTION}

\bigskip
\centerline{\it Dynamical Stability Limit}
\medskip

To obtain the dynamical stability limit using equation~(3.1), we need
to evaluate the second derivatives of the energy function with respect to
$\{\alpha_i\}=\{r,\rho_c,\lone,\ltwo,\rho_c',\lone',\ltwo'\}$ while
holding $J,~\cc$ and $\cc'$ fixed. We adopt the notation
of Appendix~A of LRS1, where many of the useful
algebraic expressions are given. Using the energy function~(2.29)
we obtain the first derivatives of the energy function as
$$\eqalign{
{\partial E\over \partial r} &= -\mu r\Omega^2+{GMM'\over r^2}
+{3GM'\delta I\over 2r^4}+{3GM\delta I'\over 2r^4},	\cr
{\partial E\over \partial \rho_c} &=
{1\over n\rho_c}U+{1\over 3\rho_c}W+{2\over 3\rho_c}T_s
+{1\over 3\rho_c}{GM'\delta I\over r^3},	\cr
{\partial E\over \partial \lone} &=
{h_{+(1)}\over \lone}T_{+}+{h_{-(1)}\over \lone}T_{-}
+{\Ich_{(1)} \over \lone}W+
{GM'\over 2 r^3\lone}(4I_{11}+I_{22}+I_{33}),\cr
{\partial E\over \partial \ltwo} &=
{h_{+(2)}\over \ltwo}T_{+}+{h_{-(2)}\over \ltwo}T_{-}
+{\Ich_{(2)} \over \ltwo}W-
{GM'\over 2 r^3\ltwo}(2I_{11}+2I_{22}-I_{33}),\cr
}\eqno(B1)$$
where we have defined $\delta I\equiv 2I_{11}-I_{22}-I_{33}$ and similarly for
$\delta I'$. The other first derivatives
$\partial E/\partial \rho_c'$, $\partial E/\partial\lone'$, and
$\partial E/\partial\ltwo'$ can be obtained from
$\partial E/\partial \rho_c$, $\partial E/\partial\lone$, and
$\partial E/\partial\ltwo$, respectively, by interchanging the
unprimed quantities and the primed quantities. In equations~(B1),
we write $T_{\pm}$ and $T_{\pm}'$ as (cf. eq.~[2.32])
$$\eqalign{
T_{\pm} &=h_{\pm}{(J_s\pm\cc)^2\over 4I_s},	\cr
T_{\pm}'&=h_{\pm}'{(J-\mu r^2\Omega-J_s\pm\cc')^2\over 4I_s'}.	\cr
}\eqno(B2)$$
Note that both $\Omega$ and $J_s$ are functions of
the adopted variables $\{\alpha_i\}$ and
the conserved quantities $J,~\cc,~\cc'$.
The expression for $\Omega(\alpha_i;J,\cc,\cc')$ is given by equation~(2.30);
the expression for $J_s$ can be obtained from equation~(2.21) as
$$J_s(\alpha_i;J,\cc,\cc')={2\over h_{+}+h_{-}}I_s\Omega(\alpha_i;J,\cc,\cc')
-{h_{+}-h_{-}\over h_{+}+h_{-}}\cc.\eqno(B3)$$

To calculate the second derivatives of the energy function, we
first calculate $(\partial\Omega/\partial\alpha_i)$ and
$(\partial J_s/\partial\alpha_i)$, then we use
$$\left({\partial^2E\over\partial\alpha_i\partial\alpha_j}\right)_{J,\cc,\cc'}
=\left[{\partial\over\partial\alpha_i}\left({\partial E\over\partial\alpha_j}
\right)\right]_{\Omega,J_s}
+\left[{\partial\over\partial\Omega}\left({\partial E\over\partial\alpha_j}
\right)\right]\left({\partial\Omega\over\partial\alpha_i}\right)
+\left[{\partial\over\partial J_s}\left({\partial E\over\partial\alpha_j}
\right)\right]\left({\partial J_s\over\partial\alpha_i}\right).
\eqno(B4)$$
For convenience, we define
$$\eqalign{
I_h &\equiv I_s/(h_{+}+h_{-}),	\cr
I_e &\equiv \mu r^2+2(I_h+I_h'),	\cr
F_d &\equiv -1+2I_h/I_e,	\cr
I_c &\equiv I_hI_h'/I_e, 	\cr \Sigma_{i} &\equiv h_{+(i)}\,
(\Omega+\Lambda)+h_{-(i)}\,(\Omega-\Lambda),~~~i=1,2, 	\cr
}\eqno(B5)$$
and similarly for $I_h'$, $F_d'$ and $\Sigma_i'$.
We obtain the following expressions:
$$\eqalign{
r^2\left({\partial^2 E\over \partial r^2}\right) &=
-\mu r^2\Omega^2-{2GMM'\over r}-{6GM'\delta I\over r^3}-{6GM\delta I'\over r^3}
-(2\mu r^2\Omega)\left(-{2\mu r^2\over I_e}\Omega\right),		\cr
r\rho_c\left({\partial^2E\over\partial r\partial\rho_c}\right) &=
-{GM'\delta I\over r^3}-(2\mu r^2\Omega)
{4I_h\over 3I_e}\Omega,			\cr
r\lone \biggl({\partial^2 E\over \partial r\partial\lone}\biggr) &=
-{3GM'\over 2r^3}(4I_{11}+I_{22}+I_{33})
-(2\mu r^2\Omega){I_h\over I_e}\Sigma_1,			\cr
r\ltwo \biggl({\partial^2 E\over \partial r\partial\ltwo}\biggr) &=
{3GM'\over 2r^3}(2I_{11}+2I_{22}-I_{33})
-(2\mu r^2\Omega){I_h\over I_e}\Sigma_2,   		\cr
\rho_c^2\biggl({\partial^2 E\over \partial \rho_c^2}\biggr) &=
{1\over 3}\left[\left({1\over 3}-{1\over n}\right)W
+2\left({2\over 3}-{1\over n}\right)T_s
-\left({2\over 3}+{1\over n}\right){GM'\delta I\over r^3}\right]
+{2\Omega\over 3}{4I_h\Omega\over 3}F_d, 			\cr
\rho_c\lone\biggl({\partial^2 E\over \partial \rho_c\partial\lone}\biggr) &=
{2 \over 3}(h_{+(1)}T_{+} + h_{-(1)}T_{-})
+{\Ich_{(1)}\over 3}W-{GM'\over 3r^3}(4I_{11}+I_{22}+I_{33})
+{2\Omega\over 3}I_h\Sigma_1F_d,		\cr
\rho_c\ltwo\biggl({\partial^2 E\over \partial \rho_c\partial\ltwo}\biggr) &=
{2 \over 3}(h_{+(2)}T_{+} + h_{-(2)}T_{-})
+{\Ich_{(2)}\over 3}W+{GM'\over 3r^3}(2I_{11}+2I_{22}-I_{33})
+{2\Omega\over 3}I_h\Sigma_2F_d,		\cr
\rho_c\rho_c'\left({\partial^2 E\over \partial\rho_c\rho_c'}\right) &=
{2\Omega \over 3}\left({8I_c\Omega\over 3}\right), 			\cr
\rho_c\lone'\left({\partial^2 E\over \partial\rho_c\lone'}\right) &=
{2\Omega \over 3}(2I_c\Sigma_1'),			\cr
\rho_c\ltwo'\left({\partial^2 E\over \partial\rho_c\ltwo'}\right) &=
{2\Omega \over 3}(2I_c\Sigma_2'),                 \cr
\lone^2\biggl({\partial^2 E\over \partial\lone^2}\biggr) &=
h_{+(11)}T_{+} + h_{-(11)}T_{-}+\Ich_{(11)}W-{6GM'I_{11}\over r^3}
+{\Sigma_1\over 2}I_h\Sigma_1F_d, 		\cr
\lone\ltwo\biggl({\partial^2 E\over \partial\lone\ltwo}\biggr) &=
h_{+(12)}T_{+} +h_{-(12)}T_{-}
+\Ich_{(12)}W+{GM'\over 2r^3}(4I_{11}-2I_{22}+I_{33})
+{\Sigma_1\over 2}I_h\Sigma_2F_d,            \cr
\lone\lone'\biggl({\partial^2 E\over \partial\lone\lone'}\biggr) &=
{\Sigma_1\over 2}(2I_c\Sigma_1'),		\cr
\lone\ltwo'\biggl({\partial^2 E\over \partial\lone\ltwo'}\biggr) &=
{\Sigma_1\over 2}(2I_c\Sigma_2'),               \cr
\ltwo^2\biggl({\partial^2 E\over \partial\ltwo^2}\biggr) &=
h_{+(22)}T_{+} + h_{-(22)}T_{-}+\Ich_{(22)}W+{3GM'I_{22}\over r^3}
+{\Sigma_2\over 2}I_h\Sigma_2F_d,          \cr
\ltwo\ltwo'\biggl({\partial^2 E\over \partial\ltwo\partial\ltwo'}\biggr) &=
{\Sigma_2\over 2}(2I_c\Sigma_2').		\cr
}\eqno(B6)$$
The other independent matrix elements can be obtained by
interchanging primed quantities and unprimed quantities in the appropriate
expression given above. For example,
$r\rho_c'(\partial^2E/\partial r\partial\rho_c')$ can be directly obtained from
the expression for $r\rho_c(\partial^2E/\partial r\partial\rho_c)$
given above, giving
$$r\rho_c'\left({\partial^2E\over\partial r\partial\rho_c'}\right) =
-{GM\delta I'\over r^3}-(2\mu r^2\Omega)
{4I_h'\over 3I_e}\Omega.	\eqno(B7)$$
The expressions for $h_{\pm(i)}$, $\Ich_{(i)}$, $h_{\pm(ij)}$
and $\Ich_{(ij)}$ are given in Appendix~A of LRS1.

\bigskip
\centerline{\it Secular Stability Limit}
\medskip

To obtain the secular stability limit along a synchronized sequence
using equation~(3.1), we need to evaluate the second derivative of the
energy function with respect to $\{\alpha_i\}$ with $J$ fixed
and holding $f_R=f_R'=0$. In this case, equation~(2.53) should be used
for the kinetic energy term.

We define
$$\eqalign{
I_t &\equiv \mu r^2+I+I',	\cr
F_s &\equiv -1+I/I_t,	\cr
}\eqno(B8)$$
and similarly for $F_s'$. The second derivatives
of the energy function are then given by:
$$\eqalign{
r^2\left({\partial^2 E\over \partial r^2}\right) &=
-\mu r^2\Omega^2-{2GMM'\over r}-{6GM'\delta I\over r^3}-{6GM\delta I'\over r^3}
-(2\mu r^2\Omega)\left(-{2\mu r^2\over I_t}\Omega\right),		\cr
r\rho_c\left({\partial^2E\over\partial r\partial\rho_c}\right) &=
-{GM'\delta I\over r^3}-(2\mu r^2\Omega)
{2I\over 3I_t}\Omega,			\cr
r\lone \biggl({\partial^2 E\over \partial r\partial\lone}\biggr) &=
-{3GM'\over 2r^3}(4I_{11}+I_{22}+I_{33})
-(2\mu r^2\Omega){I\over I_t}h_{(1)}\Omega,			\cr
r\ltwo \biggl({\partial^2 E\over \partial r\partial\ltwo}\biggr) &=
{3GM'\over 2r^3}(2I_{11}+2I_{22}-I_{33})
-(2\mu r^2\Omega){I\over I_t}h_{(2)}\Omega,   		\cr
\rho_c^2\biggl({\partial^2 E\over \partial \rho_c^2}\biggr) &=
{1\over 3}\left[\left({1\over 3}-{1\over n}\right)W
+2\left({2\over 3}-{1\over n}\right)T_s
-\left({2\over 3}+{1\over n}\right){GM'\delta I\over r^3}\right]
+{2\Omega\over 3}{2I\Omega\over 3}F_s, 			\cr
\rho_c\lone\biggl({\partial^2 E\over \partial \rho_c\partial\lone}\biggr) &=
{2 \over 3}h_{(1)}T_s
+{\Ich_{(1)}\over 3}W-{GM'\over 3r^3}(4I_{11}+I_{22}+I_{33})
+{2\Omega\over 3}h_{(1)}I\Omega F_s,		\cr
\rho_c\ltwo\biggl({\partial^2 E\over \partial \rho_c\partial\ltwo}\biggr) &=
{2 \over 3}h_{(2)}T_s
+{\Ich_{(2)}\over 3}W+{GM'\over 3r^3}(2I_{11}+2I_{22}-I_{33})
+{2\Omega\over 3}h_{(2)}I\Omega F_s,		\cr
\rho_c\rho_c'\left({\partial^2 E\over \partial\rho_c\rho_c'}\right) &=
{2\Omega \over 3}{2I I'\Omega\over 3I_t}, 			\cr
\rho_c\lone'\left({\partial^2 E\over \partial\rho_c\lone'}\right) &=
{2\Omega \over 3}h_{(1)}'{I I'\Omega\over I_t},			\cr
\rho_c\ltwo'\left({\partial^2 E\over \partial\rho_c\ltwo'}\right) &=
{2\Omega \over 3}h_{(2)}'{I I'\Omega\over I_t},                 \cr
\lone^2\biggl({\partial^2 E\over \partial\lone^2}\biggr) &=
h_{(11)}T_s +\Ich_{(11)}W-{6GM'I_{11}\over r^3}
+(h_{(1)}\Omega)h_{(1)}I\Omega F_s, 		\cr
\lone\ltwo\biggl({\partial^2 E\over \partial\lone\ltwo}\biggr) &=
h_{(12)}T_s +\Ich_{(12)}W+{GM'\over 2r^3}(4I_{11}-2I_{22}+I_{33})
+(h_{(1)}\Omega)h_{(2)}I\Omega F_s,            \cr
\lone\lone'\biggl({\partial^2 E\over \partial\lone\lone'}\biggr) &=
(h_{(1)}\Omega)h_{(1)}'{I I'\Omega\over I_t},		\cr
\lone\ltwo'\biggl({\partial^2 E\over \partial\lone\ltwo'}\biggr) &=
(h_{(1)}\Omega)h_{(2)}'{I I'\Omega\over I_t},               \cr
\ltwo^2\biggl({\partial^2 E\over \partial\ltwo^2}\biggr) &=
h_{(22)}T_s +\Ich_{(22)}W+{3GM'I_{22}\over r^3}
+(h_{(2)}\Omega)h_{(2)}I\Omega F_s,          \cr
\ltwo\ltwo'\biggl({\partial^2 E\over \partial\ltwo\partial\ltwo'}\biggr) &=
(h_{(2)}\Omega)h_{(2)}'{I I'\Omega\over I_t}.		\cr
}\eqno(B9)$$
Again, the other independent matrix elements can be obtained by
interchanging primed quantities and unprimed quantities in the appropriate
expression given above.
The expressions for $h_{(i)}$ and $h_{(ij)}$ are given in Appendix~A of LRS1.

\vskip 1.5truecm
\centerline{\bf REFERENCES}
\vskip 0.3truecm

\smallskip\noindent
Abramovici, A. et al. 1992, Science, 256, 325

\smallskip\noindent
Aizenman, M.\ L. 1968, ApJ, 153, 511

\smallskip\noindent
Alexander, M.\ E. 1973, Astron. Space Sci., 23, 459


\smallskip\noindent
Bailyn, C.D. 1993, in Dynamics of Globular Clusters: a Workshop in
Honor of I.R.~King, eds.\par S.~Djorgovski \& G.~Meylan, ASP Conf.\ Series,
in press

\smallskip\noindent
Bildsten, L., \& Cutler, C. 1992, ApJ, 400, 175

\smallskip\noindent
Burrows, A., \& Liebert, J. 1993, Rev. Mod. Phys., 65, 301

\smallskip\noindent
Chandrasekhar, S. 1939, An Introduction to the Study of
Stellar Structure,  (Chicago:
The Uni-\par versity of Chicago Press).

\smallskip\noindent
Chandrasekhar, S. 1969, Ellipsoidal Figures of Equilibrium
(New Haven: Yale University
Press)\par (Ch69)

\smallskip\noindent
Chen, K, \& Leonard, P.\ J.\ T. 1993, ApJL, 411, L75

\smallskip\noindent
Clark, J.\ P.\ A., \& Eardley, D.\ M. 1977, ApJ, 251, 311


\smallskip\noindent
Counselman, C.\ C. 1973, ApJ, 180, 307

\smallskip\noindent
D'Antona, F. 1987, ApJ, 320, 653

\smallskip\noindent
Evans, C.\ R., Iben, I., \& Smarr, L. 1987, ApJ, 323, 129


\smallskip\noindent
Hachisu, I. 1986, ApJS, 62, 461

\smallskip\noindent
Hachisu, I., \& Eriguchi, Y. 1984a, Pub.~Astron.~Soc.~Japan, 36, 239

\smallskip\noindent
Hachisu, I., \& Eriguchi, Y. 1984b, Pub.~Astron.~Soc.~Japan, 36, 259

\smallskip\noindent
Hut, P. 1980, A\&A, 92, 167

\smallskip\noindent
Iben, I., \& Tutukov, A.\ V. 1984, ApJS, 54, 335

\smallskip\noindent
Jaranowski, P., \& Krolak, A. 1992, ApJ, 394, 586

\smallskip\noindent
Kochanek, C.\ S. 1992, ApJ, 398, 234

\smallskip\noindent
Kopal, Z. 1959, Close Binary Systems (London: Chapman \& Hall)


\smallskip\noindent
Lai, D., Abrahams, A.\ M., \& Shapiro, S.\ L. 1991, ApJ, 377, 612

\smallskip\noindent
Lai, D., Rasio, F.\ A., \& Shapiro, S.\ L. 1993a, ApJS, in press (LRS1)

\smallskip\noindent
Lai, D., Rasio, F.\ A., \& Shapiro, S.\ L. 1993b, ApJL, 406, L63 (LRS2)

\smallskip\noindent
Lai, D., Rasio, F.\ A., \& Shapiro, S.\ L. 1993c, ApJ, in press (LRS3)

\smallskip\noindent
Landau, L.\ D., \& Lifshitz, E.\ M. 1987, Fluid Mechanics, 2nd Ed.\ (Oxford:
Pergamon
Press)

\smallskip\noindent
Levine, A., Rappaport, S., Putney, A., Corbet, R., \& Nagase, F. 1991,
ApJ, 381, 101

\smallskip\noindent
Levine, A., Rappaport, S., Deeter, J.\ E., Boynton, P.\ E., \& Nagase, F. 1993,
ApJ, in press

\smallskip\noindent
Mateo, M., Harris, H.\ C., Nemec, J., \& Olszewski, E.\ W. 1990, AJ, 100, 469

\smallskip\noindent
Miller, B.\ D. 1974, ApJ, 187, 609

\smallskip\noindent
Narayan, R., Paczy\'nski, B., \& Piran, T. 1992, ApJL, 395, L83

\smallskip\noindent
Ostriker, J.\ P. \& Gunn, J.\ E. 1969, ApJ, 157, 1395

\smallskip\noindent
Press, W.\ H., Flannery, B.\ P., Teukolsky, S.\ A., \& Vetterling, W.\ T. 1987
Numerical Recipes:
The\par Art of Scientific Computing (Cambridge: Cambridge Univ.\ Press)

\smallskip\noindent
Rasio, F.\ A. 1993, in Proceedings of International Workshop on Evolutionary
Links
in the Zoo of\par Interacting Binaries, ed.\ F.~D'Antona, Mem.\ Soc.\ Astron.\
Ital.,
in press

\smallskip\noindent
Rasio, F.\ A., \& Shapiro, S.\ L. 1992, ApJ, 401, 226

\smallskip\noindent
Rasio, F.\ A., \& Shapiro, S.\ L. 1993, in preparation

\smallskip\noindent
Shapiro, S.\ L., \& Teukolsky, S.\ A. 1983, Black Holes, White Dwarfs, and
Neutron Stars
(New York:\par Wiley)

\smallskip\noindent
Tassoul, J.-L. 1978, Theory of Rotating Stars
(Princeton: Princeton University Press)

\smallskip\noindent
Usov, V.\ V. 1992, Nature, 357, 472

\smallskip\noindent
Wiringa, R.\ B., Fiks, V., \& Fabrocini, A. 1988, Phys.\ Rev., C38, 1010

\smallskip\noindent
Zahn, J.-P. 1977, A\&A, 57, 383

\smallskip\noindent
Zapolsky, H.\ S., and Salpeter, E.\ E. 1969, ApJ, 158, 809

\vskip 1.5truecm
\centerline{\bf FIGURE CAPTIONS}
\vskip 0.3truecm

\medskip\noindent
{\bf FIG.~1}.---
Equilibrium curves of total energy $E$, angular momentum $J$, and orbital
angular velocity $\Omega$ as a function of $r$ along selected
Darwin-Riemann sequences. All sequences have $K=K'$,
 $n=n'=1.5$, and $\fch=\fch'=0$ (synchronized spins).
Curves corresponding to different values of the mass ratio $p=M/M'$ are shown:
$p=1$ (solid lines), $p=0.8$ (dotted lines), $p=0.6$ (short-dashed lines)
and $p=0.5$ (long-dashed lines).
To obtain convergence of all the curves at large $r$, the units of energy
and angular momentum are chosen to be $E_c\equiv GMM'/(R_o+R_o')$ and
$J_c\equiv [GM^2M'^2(R_o+R_o')/(M+M')]^{1/2}$.
The quantities $E_{\infty}
=-[(3-n)/(5-n)]GM^2/R_o-[(3-n')/(5-n')]GM'^2/R_o'$
and $\Omega_K^2=G(M+M')/r^3$ have been subtracted for convenience.

\medskip\noindent
{\bf FIG.~2}.---
Equilibrium curves of total energy as a function of binary separation along
selected Darwin-Riemann sequences containing two identical components
with $n=0$. The solid line is for the synchronized sequence,
the dotted lines are for constant-circulation sequences with, from top to
bottom, $2\cc/(GM^3R_o)^{1/2}=-0.32$, $-0.2832$, $-0.16$, and~0, corresponding
to
$\Omega_s/(GM/R_o^3)^{1/2}=0.4$, $0.354$, $0.2$, and~0.
The sequence having $2\cc/(GM^3R_o)^{1/2}=-0.2832$ (slightly thicker
dotted line) has an energy minimum located precisely on the synchronized
sequence. This point (solid round dot) marks the onset of dynamical
instability along the synchronized sequence. The energy minima mark
the onset of secular instability along the synchronized sequence and
dynamical instability along the constant$-\cc$ sequences.
The thick dashed line connects the dynamical stability limits of all
constant$_\cc$ sequences. It represents the boundary of the region containing
dynamically stable equilibrium configurations in the $(E,r)$ plane.

\medskip\noindent
{\bf FIG.~3}.---
General classification of equilibrium sequences according to terminal
configurations and stability limits.
The existence and ordering of the secular stability limits (round dots),
dynamical stability limits (square dots), and Roche limits (triangles) is
shown schematically along equilibrium energy curves.
The curves terminate at the contact solution.

\medskip\noindent
{\bf FIG.~4}.---
Diagrams distinguishing the different regimes
illustrated in Fig.~3 for synchronized models with (i) $K=K'$
(low-mass white dwarfs and planets),
(ii) $R_o/R_o'=M/M'$ (low-mass main-sequence stars),
and (iii) $R_o=R_o'$ (neutron stars and brown dwarfs).
All models have $n=n'$.

\medskip\noindent
{\bf FIG.~5}.---
Same as Fig.~4, but here for irrotational configurations ($f_R=f_R'=-2$)
with (i) $K=K'$ and (ii) $R_o=R_o'$, all with $n=n'$.

\medskip\noindent
{\bf FIG.~6}.---
Same as Fig.~4, but here for synchronized Roche-Riemann configurations with
$p=M/M'<1$ (i.e., the point mass is more massive than its finite-size
companion).
In many cases, Darwin-Riemann configurations  approach Roche-Riemann
configurations as $p\rightarrow 0$ (see text).

\medskip\noindent
{\bf FIG.~7}.---
Comparison between our results for incompressible Darwin models and
those of Hachisu \& Eriguchi (1984b). Mass ratios $p=M/M'=1.5$, 5, and~10
have been considered. The square of the orbital frequency $\Omega^2$
is plotted as a function of the total angular momentum $J$ in the
system. The units are defined in eqs.~(4.1).
The solid lines show our results, the dots are from Table~1 of HE.

\medskip\noindent
{\bf FIG.~8}.---
Comparison between our results for models of main-sequence star
binaries and those of recent SPH calculations.
All models have $n=n'=1.5$ and polytropic constants such that $R_o/M=R_o'/M'$.
Mass ratios $p=1.0$, 0.75, 0.5, and 0.25 are considered.
The dashed lines show our
quasi-analytic results, the solid lines were determined using SPH.
The onset of Roche lobe overflow
 as determined by SPH is indicated by the thick vertical
line segments marked RL. The point of first contact along the SPH
 sequence with $p=1.0$ is indicated by the thin vertical segment marked C.

\medskip\noindent
{\bf FIG.~9}.---
Comparison between our results and those of recent SPH calculations
for models with $n=n'=0.5$, $K=K'$, and $p=M/M'=0.85$.
Conventions are as in Fig.~8.
The dynamical stability limit determined by SPH is indicated by the
short vertical dashed line. The thin dashed line to the right
shows the corresponding result for a binary model containing two
rigid spheres (cf.\ \S6.3).

\medskip\noindent
{\bf FIG.~10}.---
Equilibrium curves of total angular momentum and energy for constant$-J$
sequences of Darwin-Riemann configurations with $n=n'=1.5$ and $M=M'$
In (a), the short-dashed lines are for $J/(GM^3R_o)^{1/2}=1.34$,
the dotted lines for $J/(GM^3R_o)^{1/2}=1.3202$, and the long-dashed lines
for $J/(GM^3R_o)^{1/2}=1.30$. The solid lines show the synchronized sequence
for comparison. The regions around points~B and~C are magnified in the
inserts. In (b), the dashed lines correspond to $J/(GM^3R_o)^{1/2}=1.323$.

\medskip\noindent
{\bf FIG.~11}.---
Same as Fig.~10, but for binaries with $n=n'=0$ and $M=M'$.
The short-dashed lines are for $J/(GM^3R_o)^{1/2}=1.6$, the dotted lines for
$J/(GM^3R_o)^{1/2}=1.523$, and the long-dashed lines for
$J/(GM^3R_o)^{1/2}=1.45$. The solid lines show the synchronized sequence.
The  thick dashed lines indicate the dynamical stability limit.

\medskip\noindent
{\bf FIG.~12}.---
Equilibrium curves of total angular momentum and energy for constant$-J$
sequences of Roche-Riemann configurations with $n=0$ and $p=M/M'=10$.
The short-dashed lines are for $J/(GM^3R_o)^{1/2}=0.25$, the dotted lines for
$J/(GM^3R_o)^{1/2}=0.2438$, and the long-dashed lines for
$J/(GM^3R_o)^{1/2}=0.25$. The solid lines correspond to the synchronized
sequence. The thicker lines show our ellipsoidal results, while
the lighter lines show the results obtained when the star is modeled
as a rigid sphere (see text).

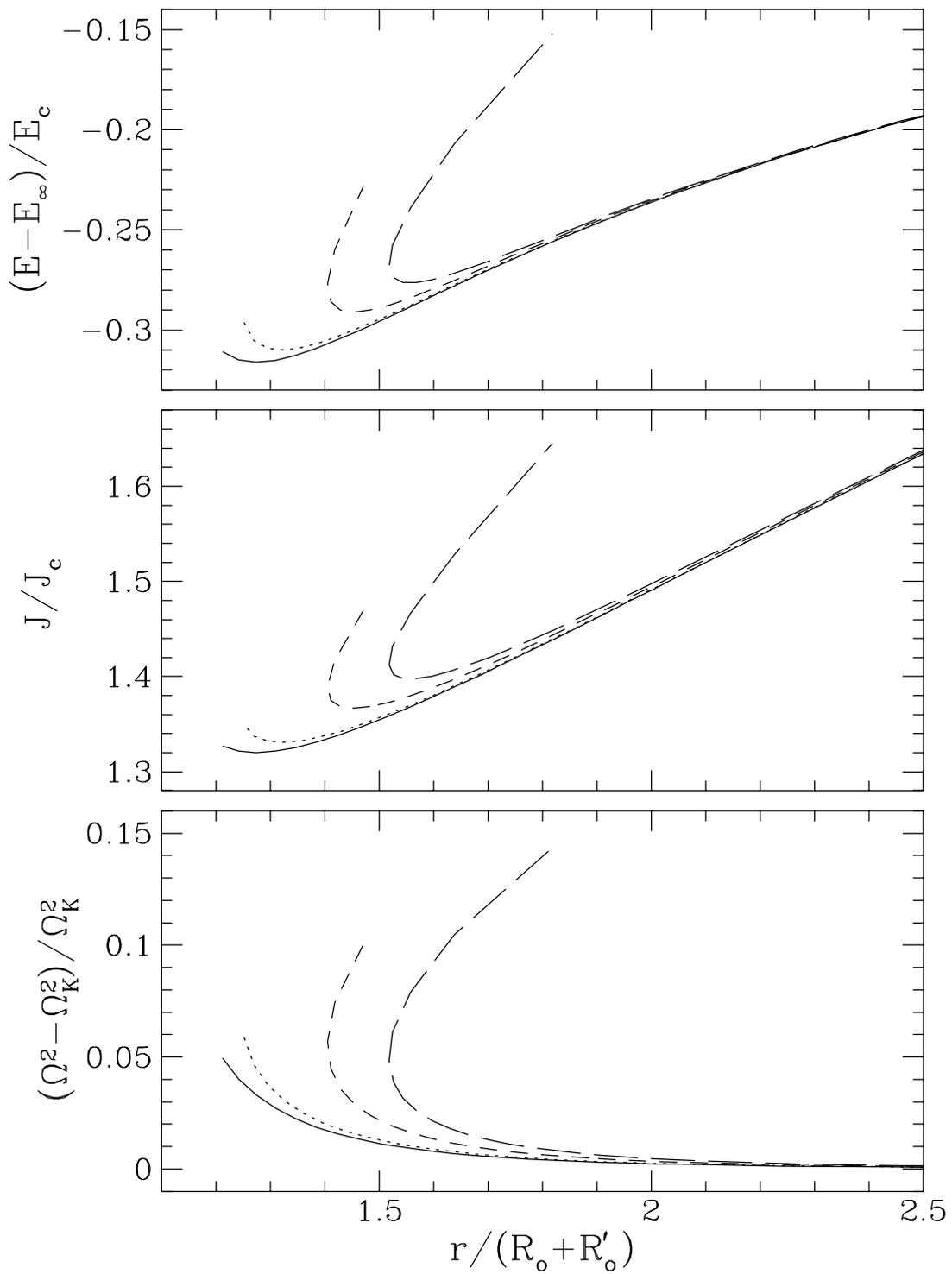
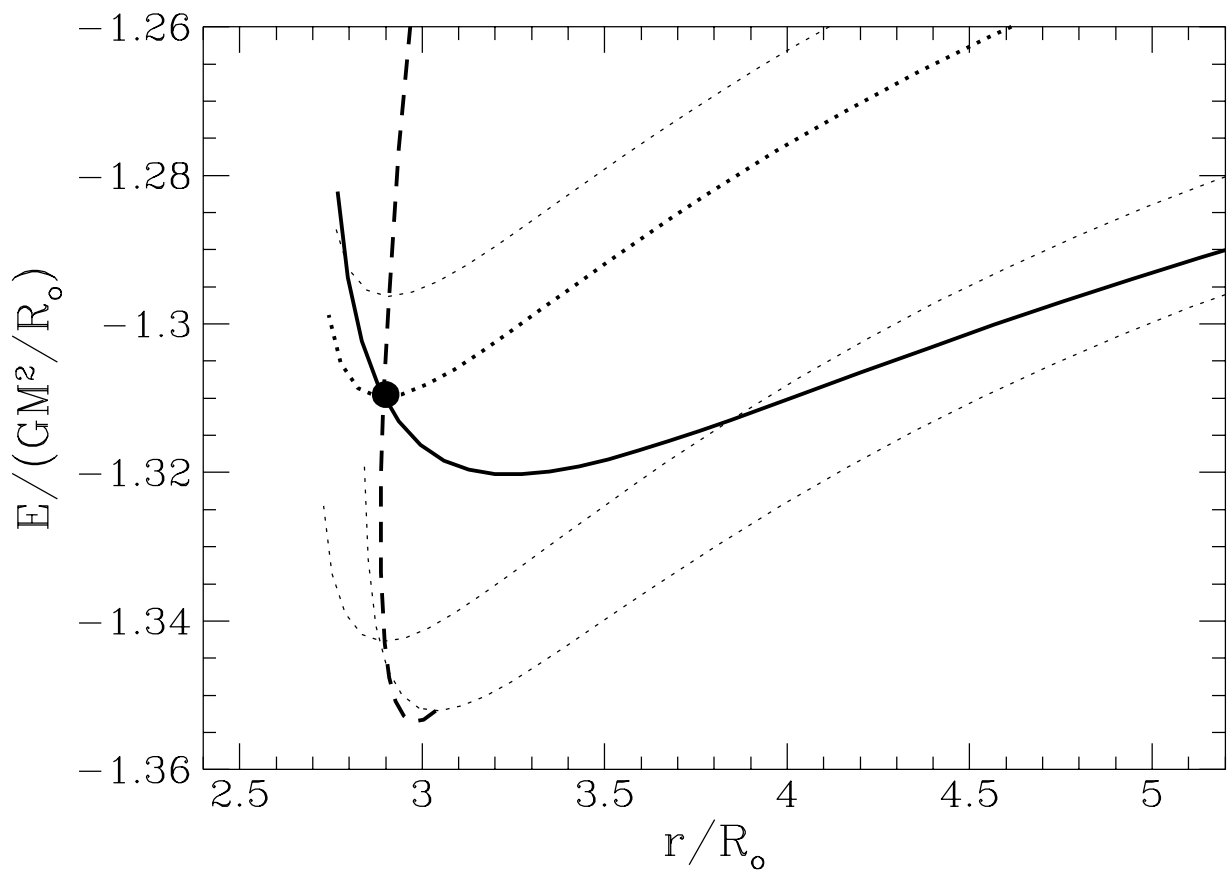
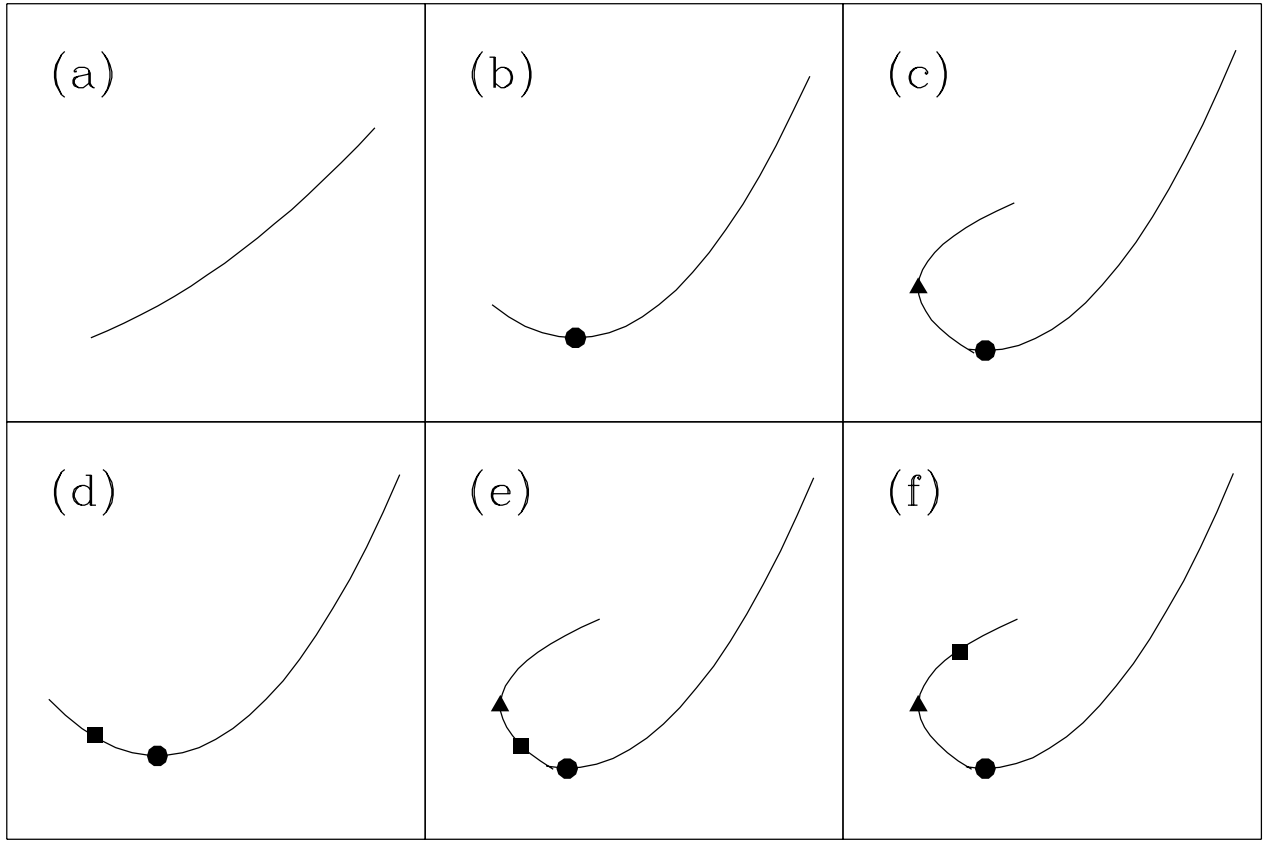
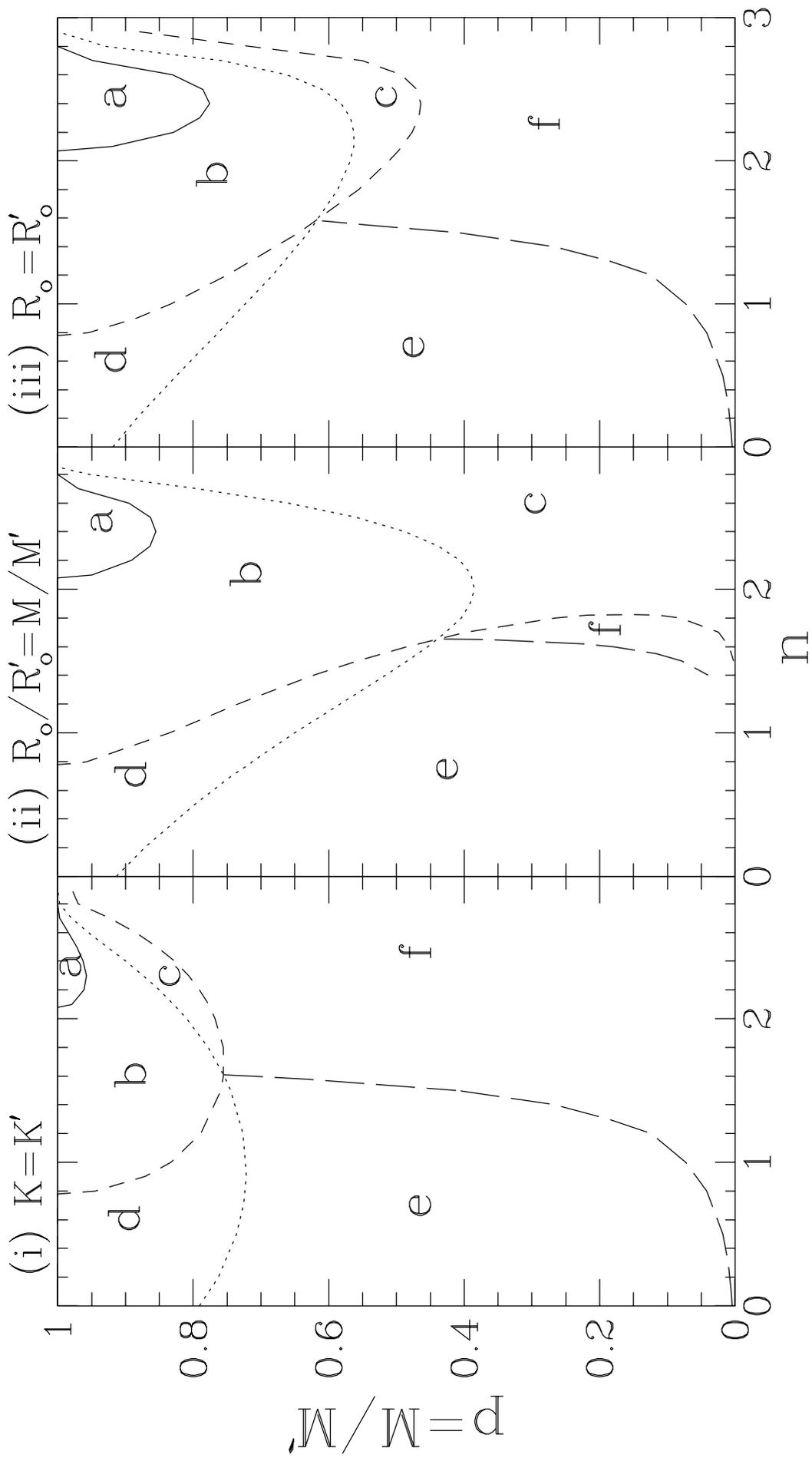
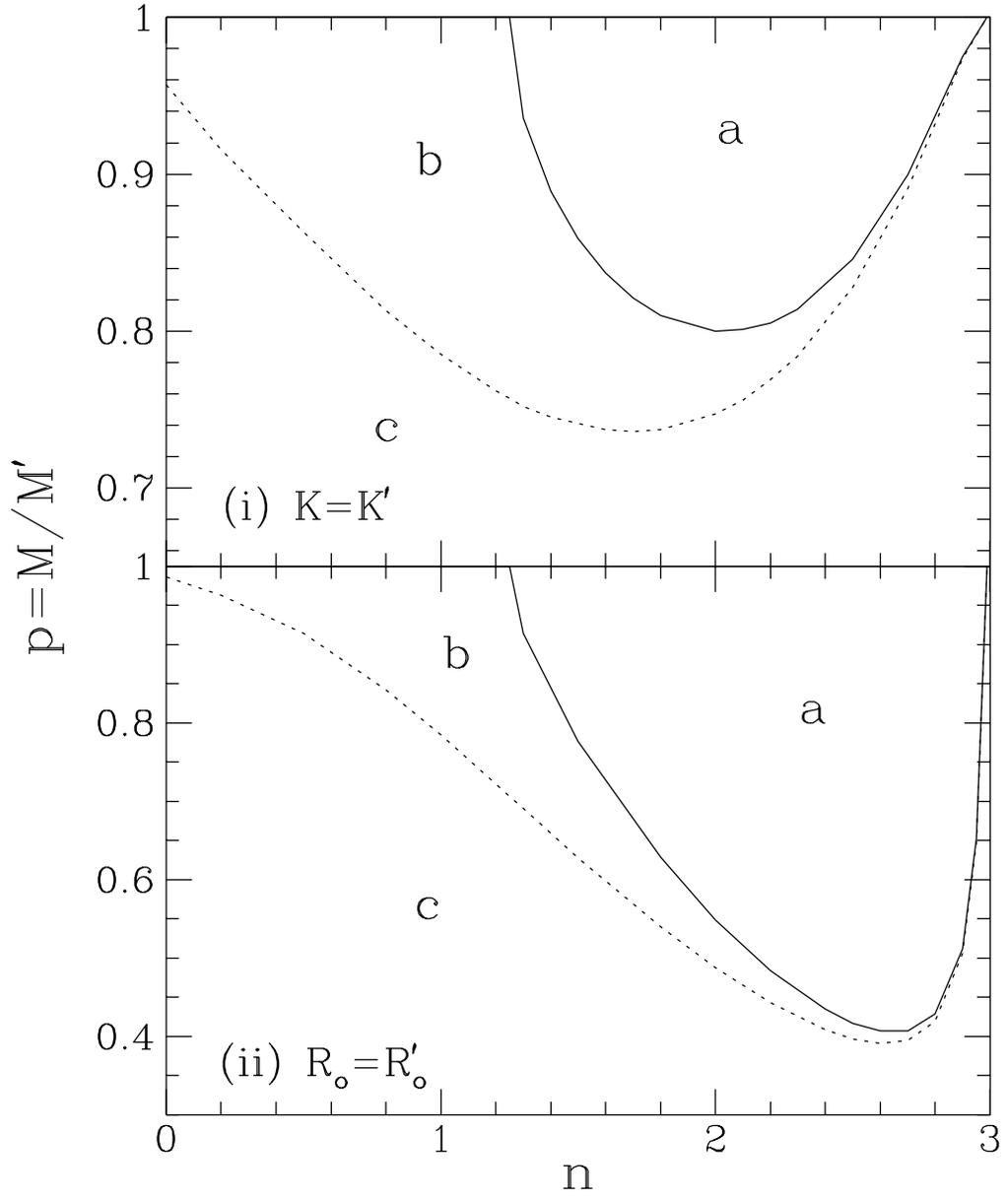
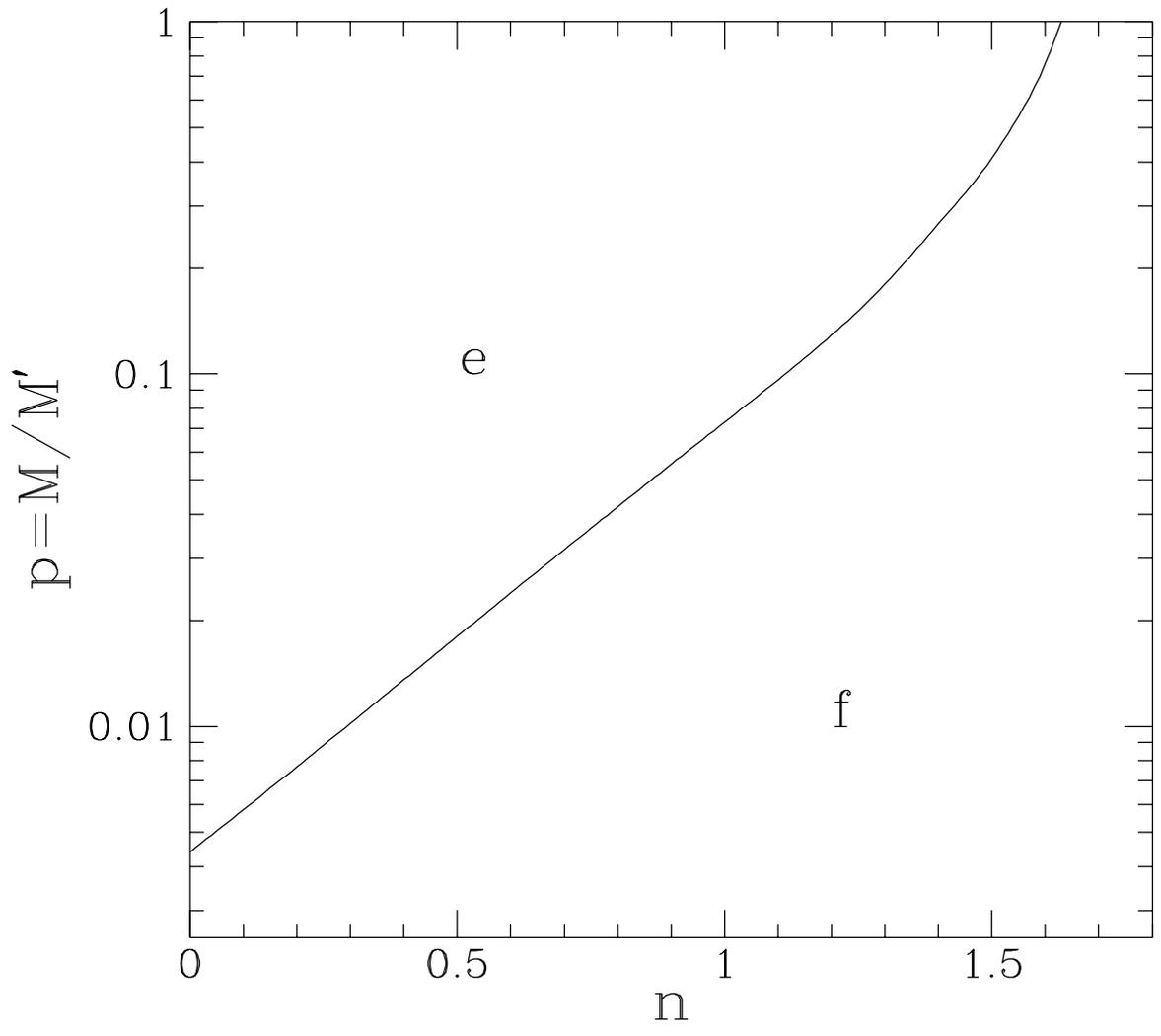
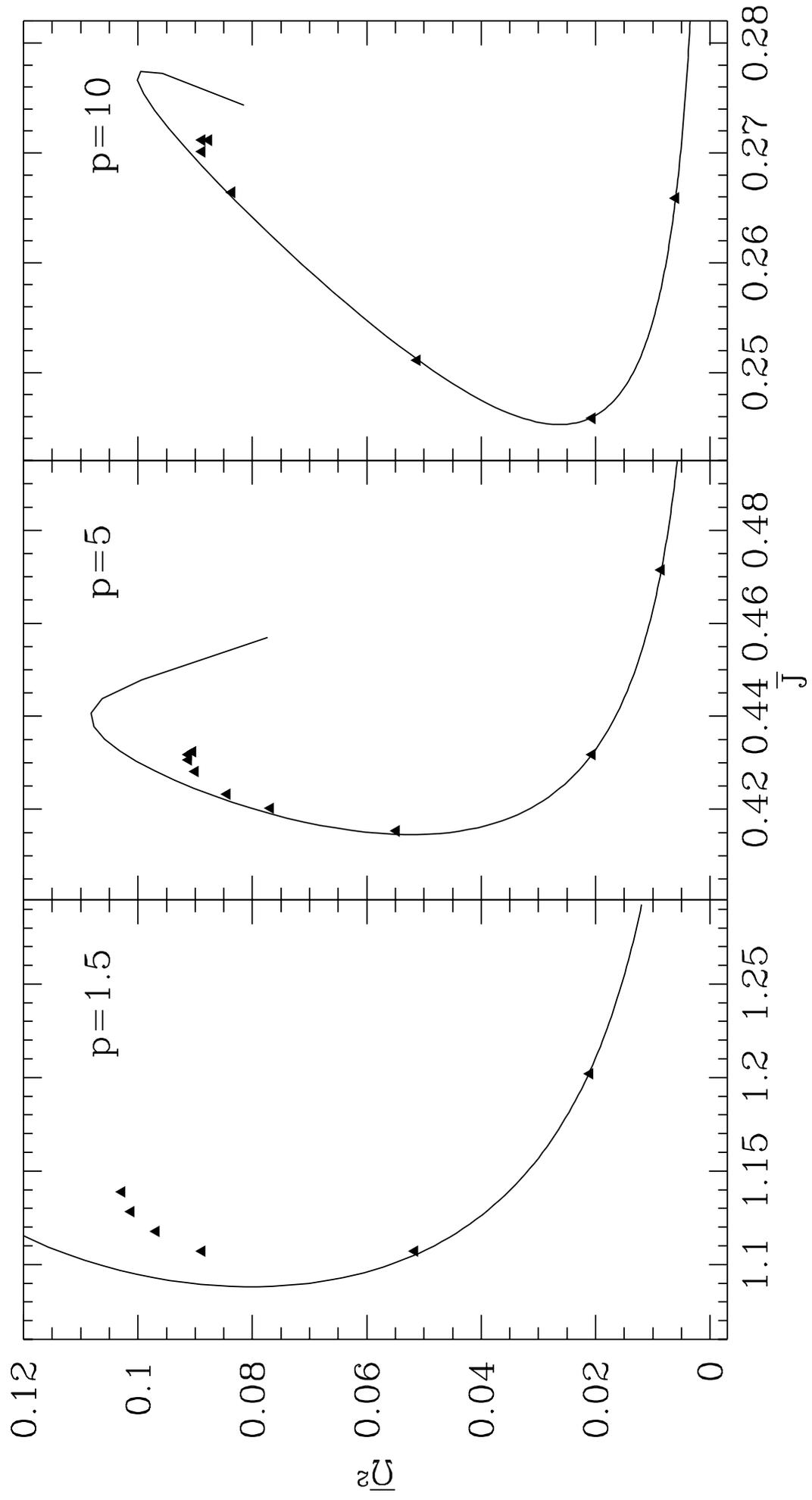
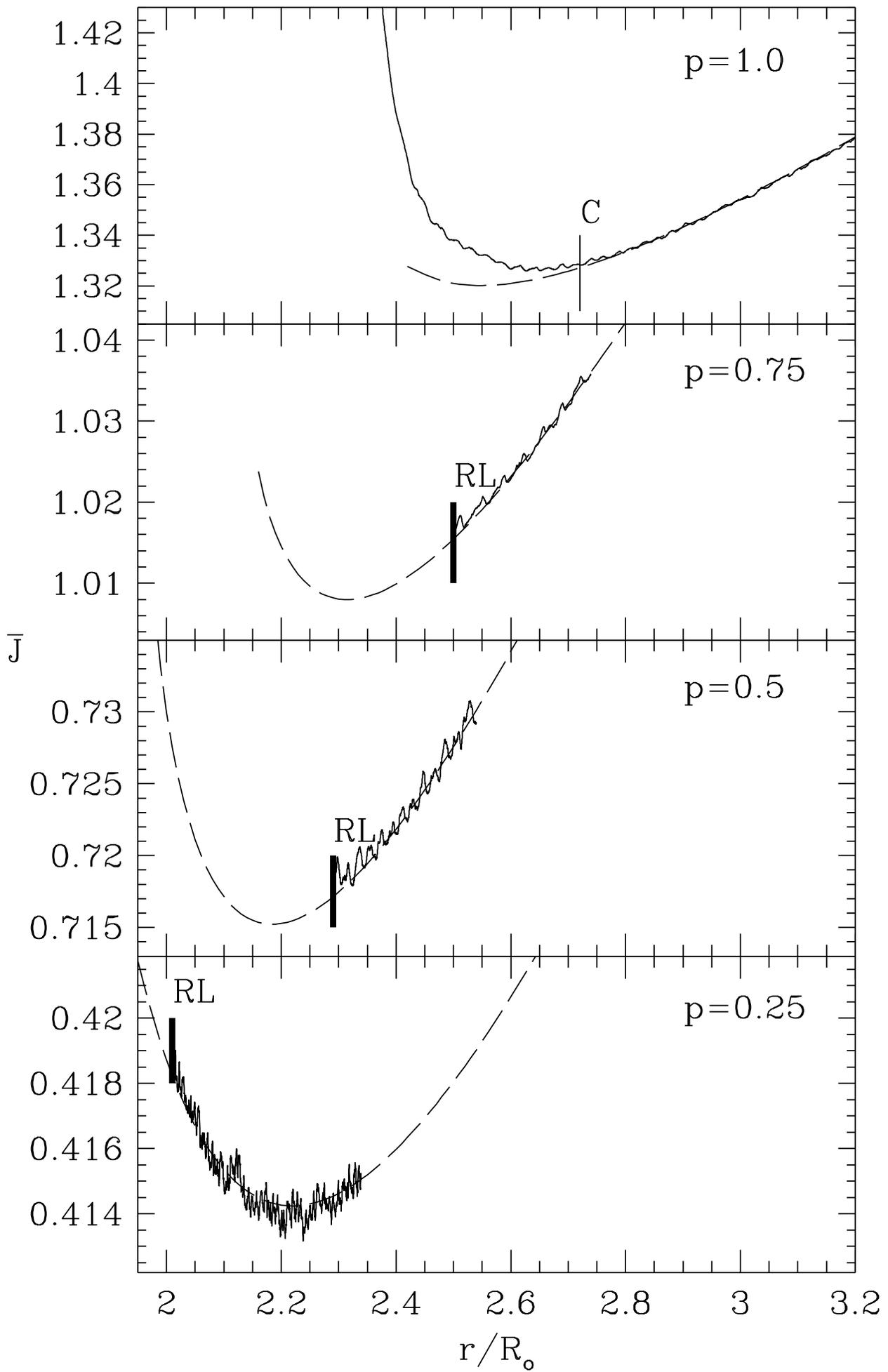
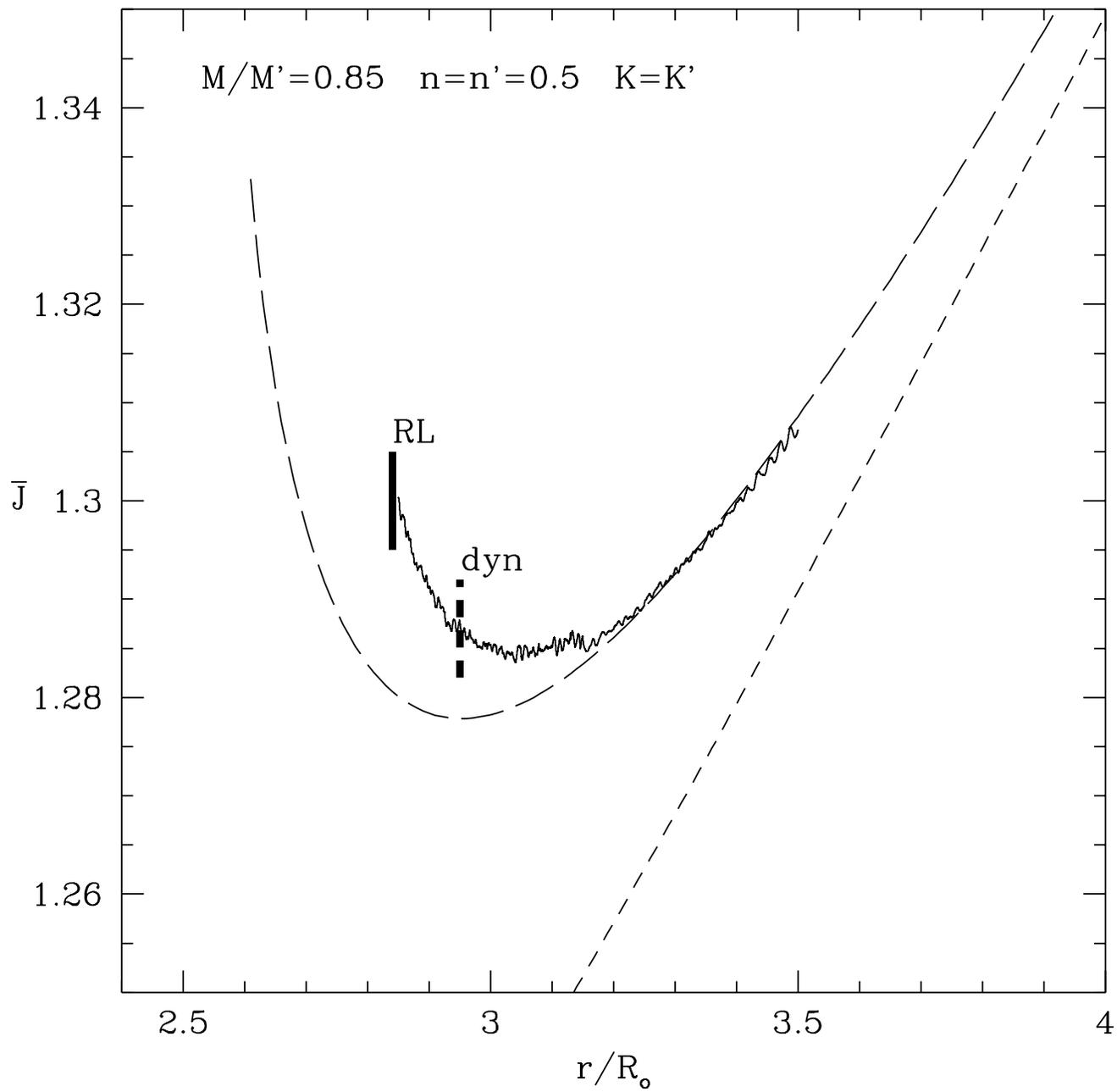
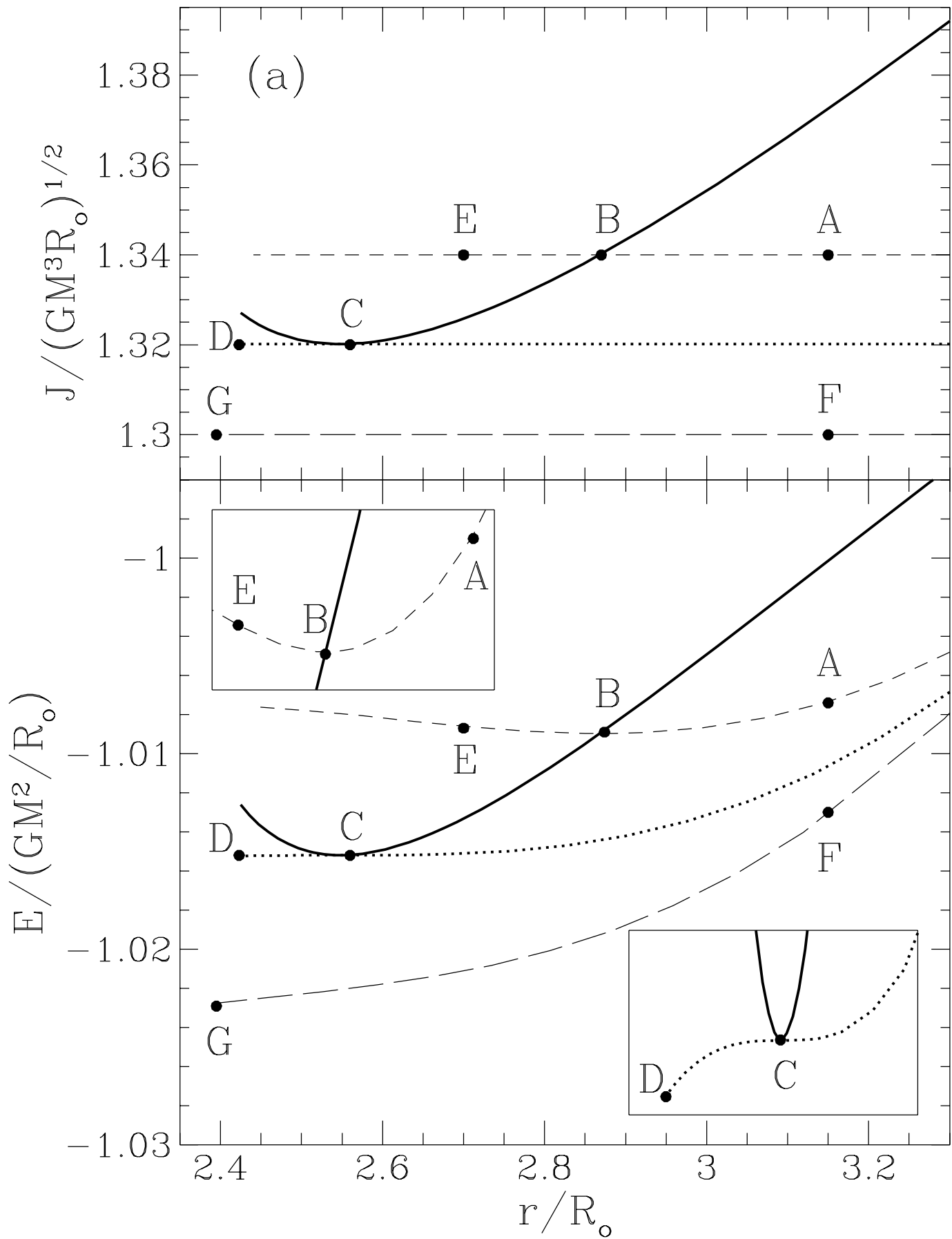
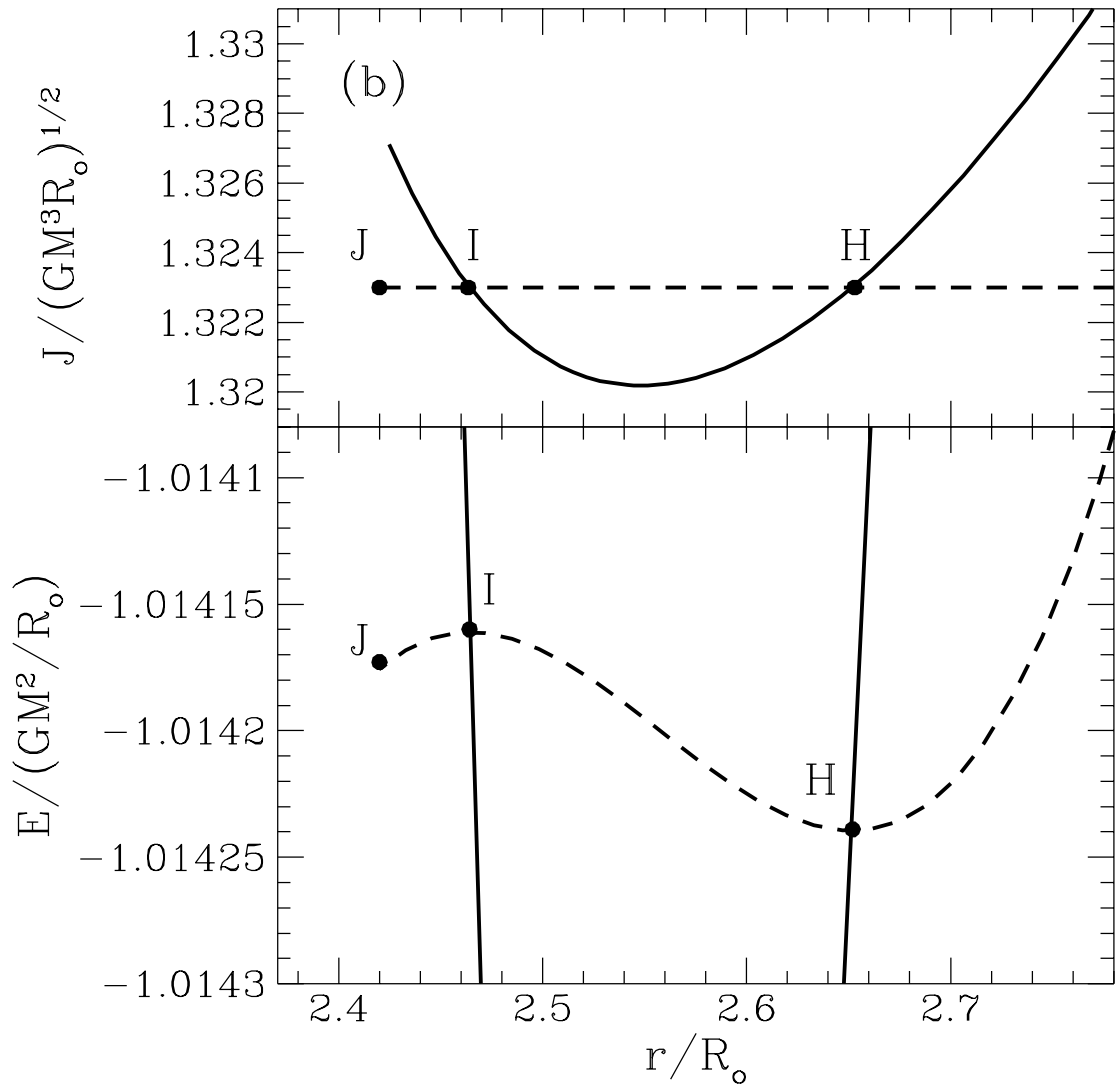
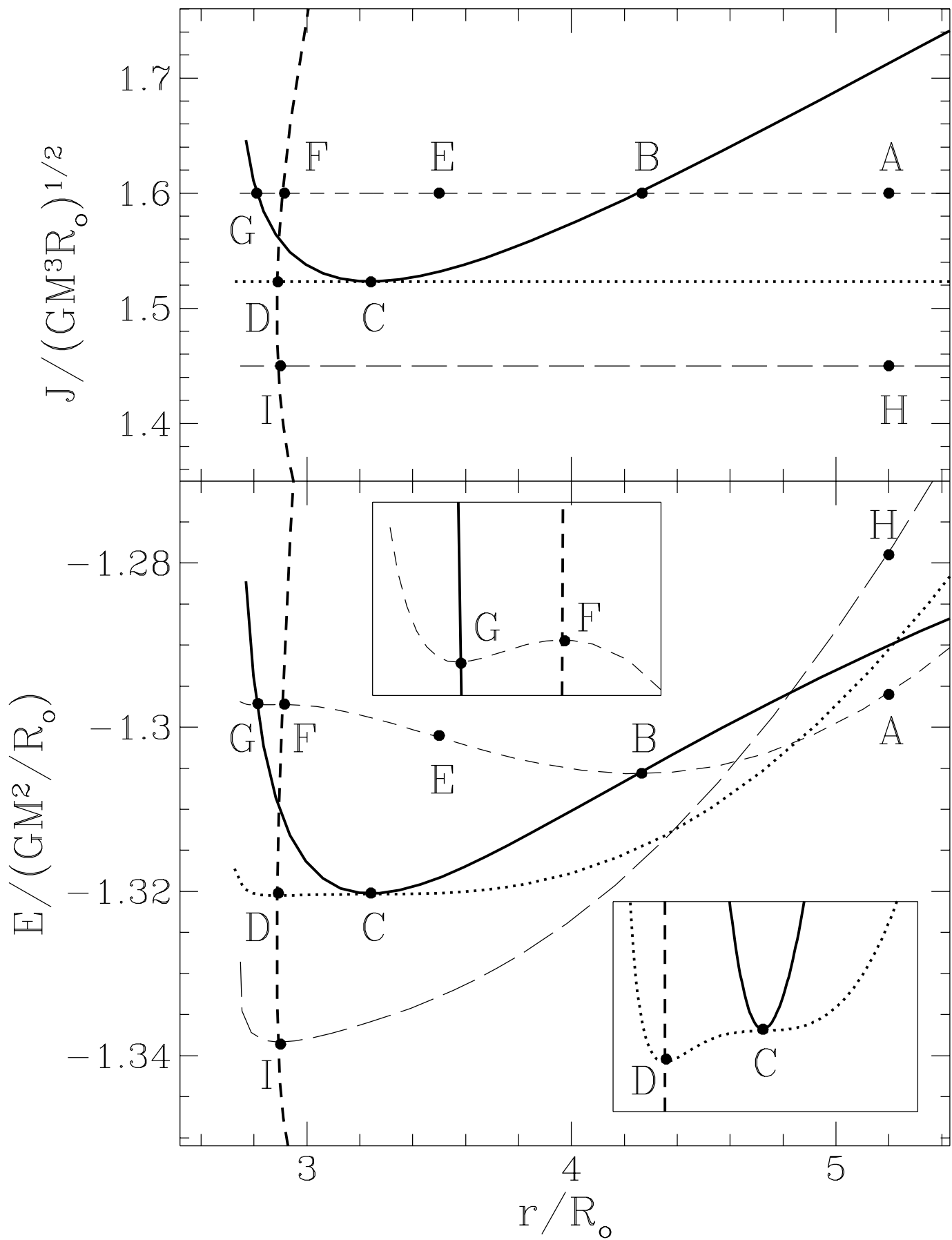
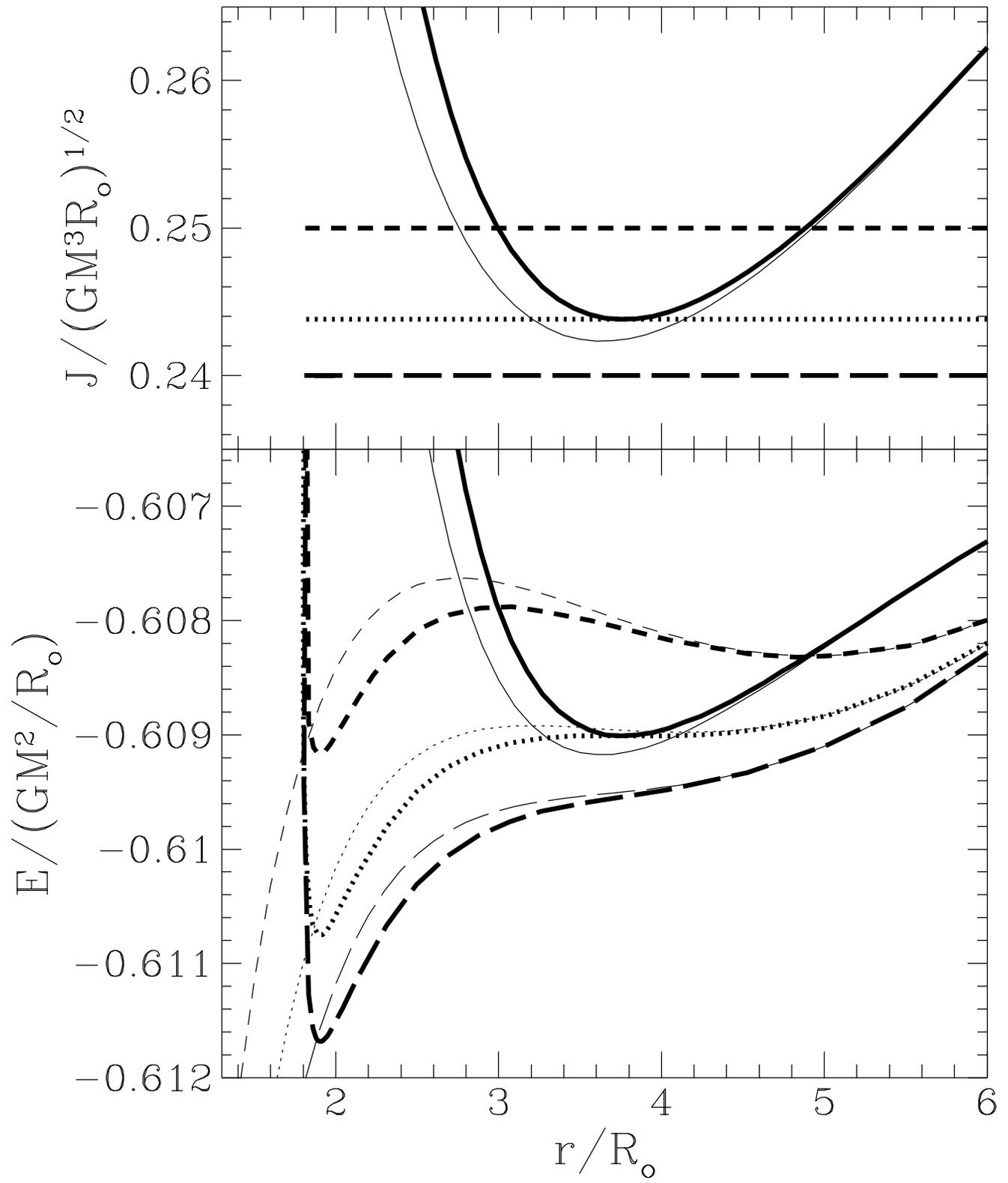
\end